\documentclass[aps, pra, reprint,preprintnumbers,amsmath,amssymb,superscriptaddress,nofootinbib,longbibliography]{revtex4-2}

\usepackage[utf8]{inputenc}
\usepackage[english]{babel}

\usepackage{graphicx}

\usepackage{leftidx}

\usepackage{diffcoeff}

\usepackage{amsfonts,amssymb,amsmath}
\usepackage{mathtools}

\usepackage{bm}
\usepackage{times}
\usepackage{bbm}
\usepackage{siunitx}
\usepackage{mathrsfs}

\usepackage{array}
\usepackage{tabularx}
\usepackage{multirow}

\usepackage{hyperref}
\hypersetup{colorlinks=true,linktoc=all,linkcolor=blue,breaklinks=true,citecolor=blue,urlcolor=blue}

\usepackage[normalem]{ulem}
\usepackage{xspace}
\usepackage[usenames,dvipsnames]{xcolor}

\addto\captionsenglish{}

\AtBeginDocument{%
    \newwrite\bibnotes
    \def\bibnotesext{Notes.bib}
    \immediate\openout\bibnotes=\jobname\bibnotesext
    \immediate\write\bibnotes{@CONTROL{REVTEX42Control}}
    \immediate\write\bibnotes{@CONTROL{%
            apsrev42Control,author="08",editor="1",pages="1",title="0",year="1"}}
    \if@filesw
    \immediate\write\@auxout{\string\citation{apsrev42Control}}%
    \fi
}%

\newcommand*{\e}{\text{e}}
\newcommand*{\kB}{k_\text{B}}

\newcommand{\eff}{\text{eff}}
\newcommand{\m}{\text{mec}}
\newcommand{\std}{\text{std}}
\newcommand{\wc}{\text{wc}}
\renewcommand{\L}{\text{L}}
\newcommand{\R}{\text{R}}
\newcommand{\opt}{\text{opt}}
\newcommand{\las}{\text{las}}
\newcommand{\In}{\text{in}}
\newcommand{\Out}{\text{out}}

\newcommand{\Om}{\Omega_\m}
\newcommand{\dOm}{\delta\Om}
\newcommand{\gam}{\Gamma_\m}
\newcommand{\omL}{\omega_\las}
\newcommand{\omd}{\omega_d}
\newcommand{\oma}{\omega_a}
\newcommand{\omc}{\omega_c}

\newcommand{\Nm}{\bar{n}_\m}
\newcommand{\Nopt}{\bar{n}_\opt}
\newcommand{\Neff}{\bar{n}_\text{fin}}

\newcommand{\Plas}{{P}_\las}

\newcommand{\Gopt}{\Gamma_\opt}
\newcommand*{\bw}[1][\mu]{\hat{b}_{\omega,#1}}
\newcommand*{\bwL}{\bw[\L]}
\newcommand*{\bwR}{\bw[\R]}

\newcommand*{\bin}[1][]{\hat{b}_{\text{in}#1}}
\newcommand*{\binL}{\bin[,\L]}
\newcommand*{\binR}{\bin[,\R]}
\newcommand*{\bout}[1][]{\hat{b}_{\text{out}#1}}
\newcommand*{\boutL}{\bout[,\L]}
\newcommand*{\boutR}{\bout[,\R]}
\newcommand*{\aL}{\alpha_\text{las}}
\newcommand*{\ain}[1][,\mu]{\hat{a}_{\text{in}#1}}
\newcommand*{\ainL}{\ain[,\L]}
\newcommand*{\ainR}{\ain[,\R]}
\newcommand*{\aout}[1][,\mu]{\hat{a}_{\text{out}#1}}
\newcommand*{\aoutL}{\aout[,\L]}
\newcommand*{\aoutR}{\aout[,\R]}

\newcommand*{\Xin}[1][,\mu]{\hat{X}_{\text{in}#1}}
\newcommand*{\XinL}{\Xin[,\L]}
\newcommand*{\XinR}{\Xin[,\R]}
\newcommand*{\Xout}[1][,\mu]{\hat{X}_{\text{out}#1}}
\newcommand*{\XoutL}{\Xout[,\L]}
\newcommand*{\XoutR}{\Xout[,\R]}

\newcommand*{\Pin}[1][,\mu]{\hat{P}_{\text{in}#1}}
\newcommand*{\PinL}{\Pin[,\L]}
\newcommand*{\PinR}{\Pin[,\R]}
\newcommand*{\Pout}[1][,\mu]{\hat{P}_{\text{out}#1}}
\newcommand*{\PoutL}{\Pout[,\L]}
\newcommand*{\PoutR}{\Pout[,\R]}

\newcommand*{\chieff}{\chi_{\m,\eff}}

\newcommand{\dXa}{\delta\hat{X}_a}
\newcommand{\dPa}{\delta\hat{P}_a}
\newcommand{\dXd}{\delta\hat{X}_d}
\newcommand{\dPd}{\delta\hat{P}_d}
\newcommand{\dQ}{\delta\hat{Q}}
\newcommand{\da}{\delta\hat{a}}
\newcommand{\dc}{\delta\hat{c}}
\newcommand{\dq}{\delta\hat{q}}
\newcommand{\deltp}{\delta\hat{p}}
\newcommand{\dd}{\delta\hat{d}}

\newcommand*{\kc}{\kappa_c}
\newcommand*{\ka}{\kappa_a}
\newcommand*{\kd}{\kappa_d}
\newcommand*{\km}{\kappa_-}
\newcommand*{\ga}{\gamma_a}
\newcommand*{\G}{\mathcal{G}}

\newcommand*{\gw}[1][c]{g^\omega_{#1}}
\newcommand*{\gk}[1][c]{g^\kappa_{#1}}
\newcommand*{\gwa}{\gw[a]}
\newcommand*{\gwd}{\gw[d]}
\newcommand*{\gka}{\gk[a]}
\newcommand*{\gkd}{\gk[d]}
\newcommand*{\gadm}{g^{\kappa,\text{asym}}}
\newcommand*{\gadp}{g^{\kappa,\text{sym}}}

\newcommand*{\tg}[1][c]{\tilde{g}_{#1}}
\newcommand*{\tga}{\tg[a]}
\newcommand*{\tgd}{\tg[d]}
\newcommand*{\tgm}[1][c]{\tilde{g}_{\m,#1}}
\newcommand*{\tgma}{\tgm[a]}
\newcommand*{\tgmd}{\tgm[d]}

\newcommand*{\bq}{\bar{q}}
\newcommand*{\bd}{\bar{d}}
\newcommand*{\ba}{\bar{a}}
\newcommand*{\bc}{\bar{c}}

\newcommand*{\tk}[1][c]{\tilde{\kappa}_{#1}}
\newcommand*{\tkd}{\tk[d]}
\newcommand*{\tka}{\tk[a]}
\newcommand*{\tkm}{\tk[-]}
\newcommand*{\tG}{\tilde{\G}}

\newcommand*{\tDc}{\tilde{\scalebox{0.85}{\ensuremath \Delta}}_c}
\newcommand*{\tDa}{\tilde{\scalebox{0.85}{\ensuremath \Delta}}_a}
\newcommand*{\tDd}{\tilde{\scalebox{0.85}{\ensuremath \Delta}}_d}
\newcommand*{\tDm}{\tilde{\scalebox{0.85}{\ensuremath \Delta}}_-}
\newcommand*{\tDp}{\tilde{\scalebox{0.85}{\ensuremath \Delta}}_+}
\newcommand*{\tOm}{\tilde{\Omega}_-}
\newcommand*{\tOp}{\tilde{\Omega}_+}
\newcommand*{\Dtot}{\bar\Delta}
\newcommand*{\ktot}{\bar\kappa}
\newcommand*{\kdiff}{\delta_\kappa}
\newcommand*{\Ddiff}{\delta_\Delta}

\DeclarePairedDelimiter{\abs}{\lvert}{\rvert}
\DeclarePairedDelimiter{\mean}{\langle}{\rangle}

\newcommand{\subfigref}[2]{\ref{#1}\hyperref[#1]{(#2)}}

\begin{document}

\title{Dissipative and dispersive cavity optomechanics with a frequency-dependent mirror}

\author{Juliette Monsel}
\affiliation{Department of Microtechnology and Nanoscience (MC2), Chalmers University of Technology, S-412 96 G\"oteborg, Sweden\looseness=-1}
\author{Anastasiia Ciers}
\affiliation{Department of Microtechnology and Nanoscience (MC2), Chalmers University of Technology, S-412 96 G\"oteborg, Sweden\looseness=-1}
\author{Sushanth Kini Manjeshwar}
\affiliation{Department of Microtechnology and Nanoscience (MC2), Chalmers University of Technology, S-412 96 G\"oteborg, Sweden\looseness=-1}
\author{Witlef Wieczorek}
\affiliation{Department of Microtechnology and Nanoscience (MC2), Chalmers University of Technology, S-412 96 G\"oteborg, Sweden\looseness=-1}

\author{Janine Splettstoesser}
\affiliation{Department of Microtechnology and Nanoscience (MC2), Chalmers University of Technology, S-412 96 G\"oteborg, Sweden\looseness=-1}

\date{\today}

\begin{abstract}
An optomechanical microcavity can considerably enhance the interaction between light and mechanical motion by confining light to a sub-wavelength volume. However, this comes at the cost of an increased optical loss rate. Therefore, microcavity-based optomechanical systems are placed in the unresolved-sideband regime, preventing sideband-based ground-state cooling. A pathway to reduce optical loss in such systems is to engineer the cavity mirrors, i.e., the optical modes that interact with the mechanical resonator. In our work, we analyze such an optomechanical system, whereby one of the mirrors is strongly frequency-dependent, i.e., a suspended Fano mirror. This optomechanical system consists of two optical modes that couple to the motion of the suspended Fano mirror. We formulate a quantum-coupled-mode description that includes both the standard dispersive optomechanical coupling as well as dissipative coupling. We solve the Langevin equations of the system dynamics in the linear regime showing that ground-state cooling from room temperature can be achieved even if the cavity is \textit{per se} not in the resolved-sideband regime, but achieves effective sideband resolution through strong optical mode coupling. Importantly, we find that the cavity output spectrum needs to be properly analyzed with respect to the effective laser detuning to infer the phonon occupation of the mechanical resonator. Our work also predicts how to reach the regime of nonlinear quantum optomechanics in a Fano-based microcavity by engineering the properties of the Fano mirror.
\end{abstract}

\maketitle

\section{Introduction}

Cavity optomechanical systems~\cite{Aspelmeyer2014Dec} find applications in quantum technologies~\cite{Barzanjeh2022Jan} and in the exploration of foundational questions~\cite{Marshall2003Sep}.
A pertinent regime in cavity optomechanics is sideband resolution, which enables manipulating quantum states of mechanical motion, for example, realizing ground-state cooling of mechanical modes~\cite{Chan2011Oct, Teufel2011Jul}, optomechanical state swapping~\cite{Palomaki2013Mar}, or non-classical mechanical state generation~\cite{Wollman2015Aug, Kotler2021May,MercierdeLepinay2021May}.
Sideband resolution has been achieved in a variety of optomechanical systems, such as Fabry-Pérot based cavities~\cite{groblacher_demonstration_2009}, microtoroids~\cite{Verhagen2012Feb}, or optomechanical crystals~\cite{Chan2011Oct}.
Commonly these setups exploit the coupling of a single optical mode to a mechanical resonator.

A major challenge in the field is to combine sideband resolution with large optomechanical coupling on the single-photon level~\cite{Rabl2011Aug, Nunnenkamp2011Aug}, where state-of-the-art systems fall two orders of magnitude short to this regime~\cite{Chan2011Oct, Teufel2011Mar}.
Recently, the concept of strongly frequency-dependent mirrors was introduced in cavity optomechanics~\cite{Naesby2018Nov,Cernotik2019Jun}.
It was shown theoretically that frequency-dependent mirrors can reduce the optical linewidth and thereby enable ground-state cooling~\cite{Cernotik2019Jun, Monsel2021Jun} or even strong optomechanical coupling~\cite{Fitzgerald2021Feb}.
Such frequency-dependent mirrors rely on Fano resonances \cite{Fan2003Mar,Limonov2017Sep} and are therefore also referred to as Fano mirrors. Examples of such frequency-dependent mirrors supporting Fano resonances are  suspended photonic crystal slabs~\cite{Fan2003Mar, Zhou2014Jan, Naesby2018Nov} or atomic arrays \cite{Bettles2016Mar, Shahmoon2017Mar, Wild2018Sep}.
In a common theoretical description of optomechanics with a Fano mirror, two coupled optical modes \cite{Limonov2017Sep} interact with a mechanical resonator \cite{Cernotik2019Jun,Manjeshwar2023May}. The description of this interaction requires extension of  the canonical optomechanical approach~\cite{Aspelmeyer2014Dec,Cernotik2019Jun} and enables new capabilities for optomechanical control. Cavity optomechanics exploiting the multi-mode nature of the optical cavity field has been analyzed in the context of back-action evasion \cite{dobrindt2010theoretical}, enhancement of nonlinearities \cite{komar2013single,burgwal2023enhanced}, and when considering the special case of two coupled optical modes that each couple to a mechanical resonator for membrane-in-the-middle systems with realizations in Fabry-Pérot based cavities \cite{thompson2008strong}, optomechanical crystals \cite{paraiso2015position,burgwal2023enhanced} or microtoroids \cite{grudinin2010phonon}. Our work focuses on a Fano-mirror optomechanical systems, where a comprehensive model and analysis of such optically and optomechanically coupled system is until now missing, thereby blocking the exploitation of opportunities that this type of system offers.

\begin{figure}[t!]
    \includegraphics[width=\linewidth]{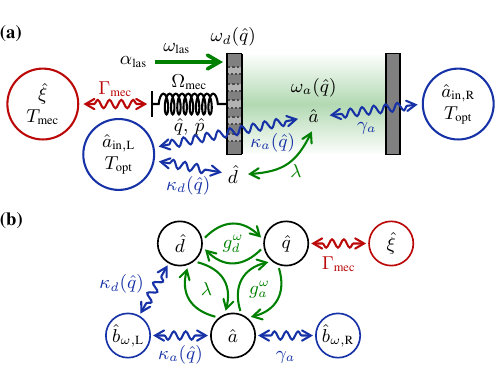}
    \caption{\label{fig:setup}
        (a) Sketch of the optomechanical setup, consisting of a double-sided optical cavity with one movable frequency-dependent mirror, and (b) the setup's coupled-mode picture.
        The cavity mode ($\hat{a}$) is coupled to electromagnetic environments on each side of the double-sided cavity (blue circles). The left mirror is frequency dependent and its internal mode ($\hat{d}$) is also coupled to the left electromagnetic environment.
        The left mirror has a mechanical degree of freedom ($\hat{q}$), which is coupled to a phonon environment (red circle). The cavity is driven with a laser at frequency $\omL$. See Secs.~\ref{sec:model:optical:micro} and~\ref{sec:model:mech:micro} for definitions of all indicated variables.
    }
\end{figure}

In our work, we provide an extensive analysis of a suspended frequency-dependent mirror coupled to an optical cavity.
We consider \emph{both} dispersive as well as dissipative optomechanical couplings~\cite{Xuereb2011Nova}, which are both observed in experiments with optomechanical microcavities~\cite{Manjeshwar2023May}.
We find that the two optical modes, i.e., the Fano mode and cavity mode, can strongly couple and will, as a result of this coupling, have a vast impact on the effective cavity decay and the optomechanical coupling. Properly choosing the parameters of this system allows access to various optomechanical regimes, including sideband resolution, single-photon strong coupling or even ultrastrong coupling, even though the optically \emph{uncoupled} system (namely in the absence of a Fano mirror) would not be able to reach these regimes. Further, the cavity output spectrum is intricate and needs careful analysis to infer the phonon occupation of the mechanical resonator. Our work extends the theory of Ref.~\cite{Cernotik2019Jun}, which considers dispersive optomechanics with a Fano mirror focusing on a specific detuning regime only, and of Ref.~\cite{Baraillon2020Sep}, which considers dispersive and two kinds of dissipative optomechanical couplings but does not address the analysis of a coupled-mirror mode.
The advantage of the presented model is that it is versatile and includes all possible mentioned couplings. It hence allows us to express the effect of the Fano mirror in terms of effective parameters in analogy to a standard optomechanical cavity system and to identify parameter regimes where ground-state cooling becomes possible thanks to the Fano-mirror coupling. We also access the intricate connection between the optical readout and the mechanical properties. As required, our general model can be reduced to a specifically chosen simpler experimental setup by simply setting irrelevant couplings to zero. It, thus, also allows comparison to the previously studied systems~\cite{Cernotik2019Jun,Baraillon2020Sep}.

The paper is structured as follows. We start with a detailed discussion of the model and its dynamics, starting from the purely optical (Sec.~\ref{sec:model:optical}) and then going to the full optomechanical model (Sec.~\ref{sec:model:mech}). In Sec.~\ref{sec:OM}, we analyze the optomechanical properties of the system and illustrate them with three sets of parameters. Then, in Sec.~\ref{sec:cooling}, we apply the insights we have gained on the system to back-action cooling, evidencing that a suitable engineering of the Fano mirror allows for ground-state cooling at room temperature. Finally, Section \ref{sec:nonlinear} details how the quantum nonlinear regime could be reached with a microcavity-based optomechanical device. We conclude in Sec.~\ref{sec:conclusion}. The appendix provides all necessary theoretical details of the approach that were left out in the main part of the paper for improved readability.

\section{Quantum coupled-mode model}\label{sec:model}

In this section, we introduce the model Hamiltonians and analyze the dynamics of an optomechanical system with a Fano mirror. We start with the purely optical part of this coupled-mode system (Fano mirror mode coupled to cavity mode) and then introduce the mechanical mode and its coupling to the optics. Compared to a previous analysis~\cite{Cernotik2019Jun}, we add the dissipative optomechanical contributions~\cite{Baraillon2020Sep, Elste2009May} to the coupled-mode model of a cavity with one Fano mirror and the optomechanical modulation of the properties of the Fano mirror.

\subsection{Optical cavity with Fano mirror}\label{sec:model:optical}

\subsubsection{Optical-modes model}\label{sec:model:optical:micro}
We study a double-sided optical cavity, as depicted in Fig.~\ref{fig:setup}, but disregard the mechanical degree of freedom at this initial stage. The cavity consists of a strongly frequency-dependent mirror (Fano mirror), such as a photonic crystal slab, on the left and a standard, highly reflective mirror on the right, inspired by Refs.~\cite{Naesby2018Nov,Manjeshwar2023May,Peralle2023Jun}. We consider a single optical cavity mode of frequency $\oma$, associated with the photon annihilation operator $\hat{a}$. In addition, the frequency dependence of the Fano mirror is modeled with another harmonic mode of frequency $\omd$ and photon annihilation operator $\hat{d}$. Due to the overlap between this guided mirror mode and the cavity mode, these two optical modes are coupled and they are described by the Hamiltonian \cite{Cernotik2019Jun}
\begin{equation}
     \hat{H}_\opt =\ \hbar\oma \hat{a}^\dagger \hat{a}^{}+ \hbar\omd \hat{d}^\dagger \hat{d}^{} + \hbar\lambda(\hat{a}^\dagger\hat{d}^{} + \hat{d}^\dagger\hat{a}^{})\label{Hopt},
\end{equation}
considering here the case of an even Fano mode\footnote{More generally, this coupling is of the form $\hbar\lambda(\hat{a}^{} + \hat{a}^\dagger)(\hat{d}^{} + \hat{d}^\dagger)$. But, depending on the symmetry of the guided mirror mode, the dominant term in the coupling is either the beam-splitter mode considered here (even mirror mode) or the two-mode-squeezing term $\hat{a}^{}\hat{d}^{} + \hat{a}^\dagger\hat{d}^\dagger$ (odd mirror mode) \cite{Cernotik2019Jun}. The analysis conducted in this article could straightforwardly be adapted to an odd mirror mode as well.}. The coupling strength $\lambda$ can be engineered via the photonic crystal or via the evanescent coupling of the two modes $a$ and $d$. The cavity mode is also coupled to the electromagnetic environments on both sides of the cavity, giving rise to the respective loss rates $\ka$ and $\ga$, while the mirror mode is only coupled to the left-hand-side environment, giving rise to the loss rate $\kd$.

\subsubsection{Langevin equations}\label{sec:model:optical:langevin}

Here, we first present the Langevin equations~\cite{Cernotik2019Jun,Gardiner1985Jun} of the optical system only, driven by a laser at frequency $\omL$, and obtain, in the frame rotating at the laser frequency,
\begin{equation}\label{Langevin optics}
    \begin{bmatrix}
        \dot{\hat{a}}\\ \dot{\hat{d}}
    \end{bmatrix}
    =
    -i\begin{bmatrix}
        \Delta_a - i(\ka + \ga) & \G\\ \G & \Delta_d - i\kd
    \end{bmatrix}
    \begin{bmatrix}
        {\hat{a}}\\ {\hat{d}}
    \end{bmatrix}
    +\text{input\ fluct}.
\end{equation}
We have introduced the total optical coupling strength, ${\G = \lambda - i\sqrt{\ka\kd}}$, including a contribution arising from the dissipative coupling of the two modes to the same environment \cite{Cernotik2019Jun}, as well as the detunings $\Delta_c = \omc - \omL$, with $c=a,d$. From the matrix characterizing the mode coupling, we obtain the complex eigenvalues
\begin{equation}\label{eq:eigenfreq}
    \Omega_\pm =\Dtot- i\ktot \pm \sqrt{\left(\Ddiff - i\kdiff\right)^2+\G^2 },
\end{equation}
which correspond to the effective resonance frequencies, $\Delta_\pm = \Re({\Omega}_\pm)$ namely the real part of $\Omega_\pm$, and their effective loss rates, $\kappa_\pm = -\Im({\Omega}_\pm)$, namely the imaginary part of $\Omega_\pm$. We have defined the average detuning $\Dtot = (\Delta_a + \Delta_d)/2$ and the average loss rate $\ktot =(\ka + \ga + \kd)/2$, as well as the difference in detunings $\Ddiff = (\Delta_a - \Delta_d)/2$ and the difference in loss rates $\kdiff =(\ka + \ga - \kd)/2$.

From the expression of the eigenvalues, Eq.~\eqref{eq:eigenfreq}, we see that we get the expected weak-coupling result for small $\G$ (small compared to the difference in frequency and loss rate between the optical modes $a$ and $d$).
However, this weak-coupling regime is not relevant here, since the dissipative part of $\G$ is $\sqrt{\ka\kd}$ which is close to the order of $\ktot$.
Furthermore, in our recent experimental realization employing a Fano mirror~\cite{Manjeshwar2023May}, it was found that $\lambda$ is larger than $\ktot$. In that case, the eigenmodes differ significantly from the cavity and mirror modes and it is rather the ``$-$'' eigenmode that gives an intuition of the behavior of the experimental system.
We will now show that we can use this optical coupling to engineer the effective optical loss rates and, in particular, make $\km$ several orders of magnitude smaller than $\kd, \ka$.
This is a prospective way to reach the (effective) resolved-sideband regime in an optomechanical system.

The effective modes discussed above are useful to interpret experimental results \cite{Manjeshwar2023May} and to identify relevant parameter regimes (Fig.~\ref{fig:regimes}).
While the identified eigenvalues $\Omega_\pm$ will occur in a number of analytical expressions in the remainder of this paper, it is in general not convenient to use the performed diagonalization in order to fully reformulate the Langevin equations of the full optomechanical system. First, we have not taken into account the optomechanical couplings of interest here during the diagonalization, so the optomechanical part of the equations would not become simpler. Second, while the operators $\hat{c}$ fulfill standard canonical commutation relations, $\left[\hat{c},\hat{c}^\dagger\right]=1$, these ``$+$'' and ``$-$'' modes do not. The reason for this is that the operators corresponding to the eigenmodes are functions of the operators $\hat{c}$ as well as of their detunings and loss rates.

\subsection{Optomechanical setup}\label{sec:model:mech}

\subsubsection{Full optomechanical model}\label{sec:model:mech:micro}

We now come to the optomechanical model, where the Fano mirror is suspended. It hence also has a mechanical degree of freedom of frequency $\Om$ and dimensionless position and momentum quadratures $\hat{q}$ and $\hat{p}$.

The overall system exhibits optomechanical effects both through dispersive and dissipative couplings, namely the respective modulations of the optical frequencies, $\oma(\hat{q})$ and $\omd(\hat{q}),$ and loss rates, $\ka(\hat{q})$ and $\kd(\hat{q})$, by the mechanical motion.
The position dependence of $\oma(\hat{q})$ comes from the change in the cavity length caused by the mechanical motion. But this change also modifies the reflection and transmission coefficients of the left mirror at the cavity resonance, affecting the loss rate $\ka(\hat{q})$, which therefore also becomes position dependent. For photonic crystals, the mechanical-position dependence of the mirror mode parameters $\omd(\hat{q})$ and $\kd(\hat{q})$ comes from the local deformation of the lattice period and air-hole radius due to the mechanical vibrations. Furthermore, in the case of microcavities~\cite{Manjeshwar2023May}, there can be an additional effect due to evanescent electromagnetic fields making the properties of the Fano mode dependent on the distance to the fixed mirror.
For most optomechanics experiments, the dependence of the optical parameters on the mechanical position $\hat{q}$ can be approximated as linear~\cite{Aspelmeyer2014Dec}, leading to linear optomechanical couplings, $\kc(\hat{q})=\kc+\sqrt{2}\gk \hat{q}$ and $\omc(\hat{q})=\omc-\sqrt{2}\gw \hat{q}$. Taking into account both optical modes $c = a,d$ and dispersive $\gw$ as well as dissipative couplings $\gk$, we define the four single-photon optomechanical couplings
\begin{align}
    \gw &= -\frac{1}{\sqrt{2}}\diffp{\omc}{q}[q=0],&\gk &= \frac{1}{\sqrt{2}}\diffp{\kc}{q}[q=0].
\end{align}
In the following $\omc$ and $\kc$ will always denote the frequencies and loss rates evaluated at zero mechanical displacement. The sign convention we have chosen for the dispersive coupling is such that the coupling strength for a standard Fabry-Pérot cavity mode is positive.
The mechanical mode and the dispersive optomechanical effects are then described by the Hamiltonian
\begin{equation}
    \hat{H}_\text{om} =  \frac{\hbar\Om}{2}(\hat{p}^2 + \hat{q}^2) - \sqrt{2}\sum_{c=a,d}\hbar \gw  \hat{c}^\dagger \hat{c}^{} \hat{q}^{}. \label{Hom}
\end{equation}

The mechanical mode is of course coupled to a phononic environment, giving rise to the damping rate $\gam$. However, in most optomechanical setups, this damping rate is orders of magnitude smaller compared to the other relevant frequencies. So, we will not consider it at the level of the Hamiltonian, but rather on the level of the Langevin equations, discussed in the next subsection.
Details about the microscopic modeling of such an environment can be found for instance in Ref.~\cite{Bowen2016}. Then, the whole setup is described by the total Hamiltonian $\hat{H}_\text{tot} = \hat{H}_\text{om} + \hat{H}_\opt + \hat{H}_\text{env} + \hat{H}_\text{int}$ \cite{Cernotik2019Jun, Elste2009May}, with the optical environment Hamiltonian $\hat{H}_\text{env} =\sum_{\mu=\L,\R} \int \dl \omega \hbar\omega \bw^\dagger \bw^{}$
and the interaction Hamiltonian for the coupling between system and optical environments
\begin{align}\nonumber
    \hat{H}_\text{int} =\,& \sum_{c} i\hbar \left[\sqrt{\frac{\kc}{\pi}}+ \frac{\gk}{\sqrt{2\pi\kc}}\hat{q}\right]\int \dl \omega (\hat{c}^\dagger \bwL^{} - \bwL^\dagger \hat{c}^{})\\
    & +  i\hbar \sqrt{\frac{\ga}{\pi}}\int \dl \omega (\hat{a}^\dagger \bwR^{} - \bwR^\dagger \hat{a}^{})\ .\label{Hint}
\end{align}
The photon annihilation operators $\bw$, with $\mu = \L,\R$, relate to the mode of frequency $\omega$ in the corresponding environment. We do not explicitly include the laser drive of frequency $\omL$ depicted in Fig.~\subfigref{fig:setup}{a} in the Hamiltonian $\hat{H}_\text{tot}$ because we model it as a part of the left environment, namely taking $\mean{\bwL} \propto \aL \delta(\omega - \omL)$, see next subsection. The amplitude of the laser drive, $\aL$, is related to the laser power by $\Plas = \hbar\omL \abs{\aL}^2$.
The expression \eqref{Hint} of the interaction Hamiltonian, $\hat{H}_\text{int}$, already contains several approximations. First, as commonly done for such systems \cite{Gardiner1985Jun, Elste2009May, Cernotik2019Jun}, we have neglected the two-mode-squeezing terms $\hat{c}^\dagger \bw^\dagger + \hat{c}^{} \bw^{}$.
Second, we have used the Markov approximation~\cite{Gardiner1985Jun}, namely we have assumed that the coupling of \textit{both} modes $a$ and $d$ is frequency independent.
Note, however, that the \emph{effective} dynamics of the cavity mode, when integrating out the Fano mirror mode, is non-Markovian~\cite{Cernotik2019Jun}. Indeed, delays and memory effects arise in the reduced evolution of mode $a$ due to its interaction with the Fano mirror since $\abs{\G}$ is not negligible compared to $\ka$. This will not be our focus though and we always consider the joint evolution of the two optical modes in the following, which is Markovian.

We have, until here, expressed all optomechanical couplings in terms of the original modes, ${c} = {a}, {d}$. However, coming back to the coupled modes introduced in Sec.~\ref{sec:model:optical}, insights about the relevant coupling regimes can be gained. The corresponding effective dispersive and dissipative optomechanical couplings are given by
\begin{equation}\label{eq:OM_eff_couplings}
    \gw[\pm] = -\frac{1}{\sqrt{2}}\diffp{\Delta_\pm}{\bq}[\bq=0],\; \gk[\pm] = \frac{1}{\sqrt{2}}\diffp{\kappa_\pm}{\bq}[\bq=0].
\end{equation}
Interestingly, we can have $\km < \Om$, even in microcavities where the optical losses are many orders of magnitude larger than the mechanical frequency.

First, in the simple case where the mirror mode and cavity mode have identical characteristics, namely $\Ddiff = \kdiff = 0$, Eq.~\eqref{eq:eigenfreq} becomes  $\Omega_\pm =\Dtot \pm \lambda - i\ktot(1 \pm \sqrt{1 - \ga/\ktot})$.
In the limit $\ga \ll \ktot$, we then find that $\km \simeq \ga/2$ and therefore the optical linewidth can be significantly reduced, even reaching the resolved sideband limit if\footnote{Note that one does not need to have perfectly identical optical modes to reach this effective resolved-sideband regime. Instead it needs to be in the regime ${\ktot \lesssim \lambda}$, $\ga < 2\Om$ and $\Ddiff^2/\ktot, \kdiff^2/\ktot \ll \Om$.} $\ga < 2\Om$. The corresponding effective dispersive and dissipative couplings then become $\gw[-] = (\gwa + \gwd)/2$ and $\gk[-] \simeq \ga(\gkd - \gka)/4\ktot$, leading in most cases to a mainly dispersive optomechanical coupling, $\gw[-] \gg \gk[-]$.

Furthermore, when $\Ddiff = \kdiff \lambda/\sqrt{\ka\kd}$, we can easily compute the real and imaginary part of Eq.~\eqref{eq:eigenfreq}, ${\Delta_\pm = \bar{\Delta} \pm \lambda\sqrt{1 + \kdiff^2/\ka\kd}}$ and $\kappa_\pm = \ktot \pm \sqrt{\ktot^2 - \ga\kd}$. In this limit and in the regime $\ga \ll \ktot$ of interest here, we have $\km \simeq \ga\kd/2\ktot$. It is hence useful to study the case where $\kd \ll \ka$ when $\ga$ cannot be made of the same order of magnitude as $\Om$.
A summary of the possible parameter regimes, focusing on the ``$-$'' effective mode, is given in Fig.~\ref{fig:regimes}.

\subsubsection{Langevin equations}\label{sec:model:mech:Langevin}

Following the usual derivation for intput-output equations \cite{Gardiner1985Jun, Cernotik2019Jun} (see details in Appendix \ref{app:Langevin}),
we obtain the Langevin equations, in the frame rotating at the laser frequency,
\begin{align}\label{Langevin eqs}
\! \dot{\hat{a}}=\,& -(i\Delta_a + \ka + \ga) \hat{a} + \sqrt{2}(i \gwa - \gka)\hat{q}\hat{a}-i\G \hat{d} \\\nonumber
&- \sqrt{2}\gadp \hat{q}\hat{d}+ \left(\sqrt{2\ka} + \frac{\gka}{\sqrt{\ka}} \hat{q}\right)\binL + \sqrt{2\ga}\binR,\\\nonumber
\! \dot{\hat{d}}=\,& -(i\Delta_d + \kd) \hat{d} +\sqrt{2}(i \gwd - \gkd)\hat{q}\hat{d} -i\G \hat{a} \\\nonumber
&- \sqrt{2}\gadp\hat{q}\hat{a} + \left(\sqrt{2\kd} + \frac{\gkd}{\sqrt{\kd}} \hat{q}\right)\binL,\\\nonumber
\! \dot{\hat{q}}=\,& \Om \hat{p},\\\nonumber
\!\dot{\hat{p}}=\,& -\Om\hat{q} -\gam \hat{p}  + i\sqrt{2}\gadm(\hat{a}^\dagger\hat{d}- \hat{d}^\dagger\hat{a}) \\\nonumber
&+ \sum_{c} \left[\sqrt{2}\gw  \hat{c}^\dagger \hat{c} - i\frac{\gk}{\sqrt{\kc}}(\hat{c}^\dagger \binL^{} - \binL^\dagger \hat{c}) \right] + \sqrt{\gam}\hat{\xi}.
\end{align}
We have denoted the effective optomechanical couplings arising from the dissipative optomechanical effects combined with the coupling of the mirror mode and cavity mode to the same (left) environment as
\begin{align}\label{g_ad}
    \gadp &= \frac{\sqrt{\ka\kd}}{2}\left(\frac{\gka}{\ka} + \frac{\gkd}{\kd} \right),\\\nonumber
    \gadm &= \frac{\sqrt{\ka\kd}}{2}\left(\frac{\gka}{\ka} - \frac{\gkd}{\kd} \right).
\end{align}
The cavity is driven through the left mirror, therefore the average input amplitudes, in $\sqrt{\text{Hz}}$, are $\mean{\binL} = \aL$ and $\mean{\binR} = 0$.

The optical output fields of interest are given by the input-output relations \cite{Gardiner1985Jun}
    \begin{align}\label{input-output}
        \boutL(t) =\,&  \binL(t) - \sum_c \left(\sqrt{2\kc} + \frac{\gk}{\sqrt{\kc}} \hat{q}(t)\right)\hat{c}(t),\\
        \boutR(t) =\,&  \binR(t) - \sqrt{2\ga}\hat{a}(t), \nonumber
    \end{align}
In the limit $\gka, \gwd, \gkd \to 0$, with only dispersive optomechanical effects on the cavity mode, Eqs.~\eqref{input-output} and \eqref{Langevin eqs}
give back the results derived in Ref.~\cite{Cernotik2019Jun}.

\subsection{Linearized dynamics}\label{sec:linearized om}

From now on, we consider that the laser drive is rather strong, such that the numbers of photons in the light fields in the cavity and in the mirror are large compared to their fluctuations. Furthermore, the optical frequencies are such that $\hbar\oma, \hbar\omd \gg \kB T_\opt$, therefore we can consider that the temperature $T_\opt$ in optical environments is zero and neglect thermal fluctuations. As a consequence, we can rewrite $\binL = \aL + \ainL$ and $\binR = \ainR$, where $\ain$ corresponds to the vacuum fluctuations and its only non-zero correlation function is
\begin{equation}
    \mean{\ain^{}(t)\ain^\dagger(t')} = \delta(t-t').\label{correlations_a_in}
\end{equation}
In contrast, concerning the mechanics, we are in the high-temperature limit, $\hbar\Om \ll \kB T_\m$. The input noise of the mechanics is determined by $\langle\hat{\xi}\rangle=0$ and by the correlation function\footnote{For simplicity, we approximate here the correlation function by its classical value, but we checked that this does not impact our results.
In general, this approximation might give inconsistencies in the quantum regime, even in the high-temperature limit, because it does not preserve the commutation relations between $\hat{q}$ and $\hat{p}$~\cite{Giovannetti2001Jan}. We discussed in particular the impact on the phonon number in the mechanics in appendix of a previous work, Ref.~\cite{Monsel2021Jun}.}
\begin{equation}
    \mean{\hat{\xi}(t)\hat{\xi}(t')} = (2\Nm + 1)\delta(t - t'),\label{correlations_xi}
\end{equation}
with the average  phonon number in the mechanical environment $\Nm = (\exp(\hbar\Om/\kB T_\m) - 1)^{-1}$.
We now split all the operators into a semi-classical average value and a fluctuation operator: $\hat{a} = \ba + \da$, $\hat{d} = \bd +\dd$, $\hat{q} = \bq + \dq$ and $\hat{p} = \bar{p} + \deltp$.

\subsubsection{Semiclassical steady state}

As a first step towards the solution of the Langevin equations~\eqref{Langevin eqs}, we here present the semiclassical steady-state solution. It is given by the expressions
\begin{align}\label{ss}
    \ba =\,& i\frac{\lambda\sqrt{2\tkd} - \tDd\sqrt{2\tka}}{\tOp \tOm}\aL,\\\nonumber
    \bd =\,& \frac{i(\lambda\sqrt{2\tka} - \tDa\sqrt{2\tkd}) - \ga\sqrt{2\tkd}}{\tOp \tOm}\aL,\\\nonumber
    \bq =\,& \frac{\sqrt{2}}{\Om}\sum_{c} \left[\gw\abs{\bc}^2 - i\frac{\gk}{\sqrt{2\kc}}\left(\bc^*\aL - \bc\aL^*\right)\right] \\\nonumber& +\frac{i\sqrt{2}}{\Om}\gadm(\ba^*\bd - \bd^*\ba),\\\nonumber
    \bar{p}  = &0\nonumber\ .
\end{align}
Here, we have introduced a tilde to indicate that parameters are taken at the average mechanical displacement $\bq$. We consider the most general situation of different optomechanical couplings. In particular, the detunings are generally impacted by dispersive optomechanical couplings, $\tDc = \Delta_c - \sqrt{2}\gw\bq$, while the loss rates ${\tk = \kc + \sqrt{2}\gk \bq}$ are impacted by dissipative optomechanical couplings. Also the effective coupling between cavity and Fano mirror ${\tG = \lambda -i \sqrt{\tka\tkd}}$ has a contribution due to the dissipative optomechanical couplings.
Note, however that these shifts in the steady-state solution due to the average mechanical displacement $\bq$ are negligibly small in typical experiments considered in the remainder of this paper. This also means that while the system of equations \eqref{ss} is nonlinear, we assume in the following that we are in a parameter regime where the displacement corrections are small and the system is stable. Eqs.~\eqref{ss} hence give a single well-defined steady state of the coupled system (see stability analysis in Appendix \ref{app:stability}).

Importantly, in Eq.~\eqref{ss}, we can clearly observe the modifications that arise with respect to a single optical cavity, where one would have $\ba=\aL\sqrt{2\ka}/[\ka + \ga +i \tDa]$, due to the coupling to the Fano mirror and due to dispersive and dissipative optomechanical couplings.
The coupling via $\tG$ to the Fano resonance modifies the effective environment coupling of the cavity as well as the effective resonance frequency.

\subsubsection{Linearized Langevin equations}

We now linearize the Langevin equations \eqref{Langevin eqs} around this semiclassical steady state \eqref{ss}, keeping terms only up to the first order in the fluctuations (see Appendix \ref{app:linearization} for a discussion of the validity of this approximation),
\begin{align}\label{Langevin lin eqs}
    \delta\dot{\hat{a}} =\,&-(i\tDa + \tka + \ga) \da +i\sqrt{2}\tga\dq -i\tG\dd  \\\nonumber&+ \sqrt{2\tka}\ainL+ \sqrt{2\ga}\ainR,\\\nonumber
    \delta\dot{\hat{d}} =\,&-(i\tDd + \tkd) \dd +i\sqrt{2}\tgd\dq  -i \tG\da + \sqrt{2\tkd}\ainL ,\\\nonumber
    \delta\dot{\hat{q}} =\,&\Om \deltp,\\\nonumber
    \delta\dot{\hat{p}}
    =\,& -\Om\dq -\gam \deltp+  \sum_{c}\sqrt{2} (\tgm \dc^\dagger +\tgm^*\dc)\nonumber\\\nonumber
    &+ c_X\sqrt{2}\XinL + c_P\sqrt{2}\PinL+ \sqrt{\gam}\hat{\xi}.
\end{align}
We have defined the effective optomechanical couplings
\begin{align}\label{effective couplings}
    \tga &=  \gwa\ba +i \gka\left(\ba  -\frac{\aL}{\sqrt{2\ka}}\right) + i\gadp\bd,\\
    \tgd &= \gwd\bd +i \gkd\left(\bd -\frac{\aL}{\sqrt{2\kd}}\right) + i\gadp\ba ,\nonumber\\
    \tgma &= \gwa\ba - i\gka\frac{ \aL}{\sqrt{2\ka}}+ i\gadm\bd,\nonumber\\
    \tgmd &= \gwd\bd - i\gkd\frac{ \aL}{\sqrt{2\kd}}-i\gadm\ba,\nonumber
\end{align}
the effective couplings of the mechanical resonator to the left optical environment
\begin{align}
    c_X & = -\sum_c\frac{\gk}{\sqrt{\kc}}\Im(\bc),&
    c_P & = \sum_c\frac{\gk}{\sqrt{\kc}}\Re(\bc),
\end{align}
and the quadratures of the optical input noises $\Xin=\frac{1}{\sqrt{2}}(\ain^\dagger+\ain)$ and $\Pin=\frac{i}{\sqrt{2}}(\ain^\dagger-\ain)$.
Each effective optomechanical coupling defined in Eqs.~\eqref{effective couplings} is divided into three parts:
the dispersive coupling contribution, the dissipative coupling contribution and a cross contribution, see Eq.~\eqref{g_ad}, due to the combination of the dissipative optomechanical effects and the coupling of the mirror mode and cavity mode to the same bath.
In the limit $\lambda,\kd \to 0$, we recover the results for a standard Fabry-Pérot cavity with both dispersive and dissipative optomechanical effects \cite{Elste2009May}.

Finally, we linearize the input-output relation Eqs.~\eqref{input-output} and get
\begin{align}\label{input-output lin}
    \aoutL(t) =\,&  \ainL(t) - \sum_c \sqrt{2\tk}\dc(t) -(c_P - i c_X) \dq(t)\nonumber,\\
    \aoutR(t) =\,&  \ainR(t) - \sqrt{2\ga}\da(t).
\end{align}

\begin{figure*}
    \includegraphics[width=\linewidth]{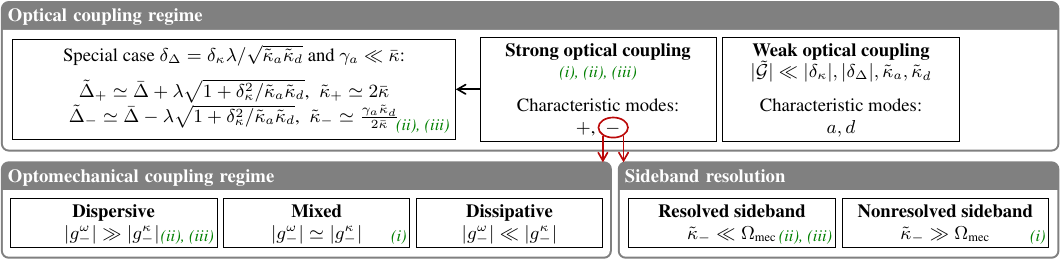}
    \caption{\label{fig:regimes}
        Illustration of the relevant parameter regimes. Each box details one aspect of the different regimes for a specific set of parameters. We have indicated in green in which regime the studied devices (see Sec.~\ref{sec:devices} and Table~\ref{tab:params}) are operating. In this article, where we focus on strong optical coupling, the effective ``$-$'' mode (highlighted in red) has a small linewidth and therefore characterizes the physics of the device, in contrast to the highly damped (and hence irrelevant) ``$+$'' mode. The optomechanical-coupling and sideband-resolution regimes in the lower two boxes therefore refer to the "$-$" mode.
    }
\end{figure*}

\subsubsection{Solution in the frequency domain}\label{sub:frequency_solution}

To determine the properties of this complex system, which will be analyzed in Sec.~\ref{sec:OM}, we solve the Langevin equations for $\da, \dd, \dq$ (Eqs.~\eqref{Langevin lin eqs}) in the frequency domain.
The full solution is given in Appendix \ref{app:solution Langevin}. Here, we present the solution for the  mechanical position fluctuations together with the relevant susceptibilities
\begin{align}\label{solution Langevin:q}
    \!\!\chieff^{-1}[\omega]\dq  =\nonumber
    & \sum_{\mu, c}\!\left(\tgm^* C^c_\mu[\omega] + \tgm C^c_\mu[-\omega]^*\right)\! \Xin\\\nonumber
    & +i\sum_{\mu, c}\!\left(\tgm^* C^c_\mu[\omega]-\tgm C^c_\mu[-\omega]^* \right)\!\Pin\\
    &+ c_X\sqrt{2}\XinL\! + c_P\sqrt{2}\PinL\!+ \sqrt{\gam}\hat{\xi}.\!\!
\end{align}
We have defined
\begin{align}\label{C_mu}
    C^a_\L[\omega] &= \frac{ i\sqrt{2\tkd}\tG - \sqrt{2\tka}\chi_d^{-1}[\omega]}{(\tOp - \omega)(\tOm - \omega)},\\\nonumber
    C^a_\R[\omega] &= \frac{-\sqrt{2\ga}\chi_d^{-1}[\omega] }{(\tOp - \omega)(\tOm - \omega)},\\\nonumber
    C^d_\L[\omega] &= \frac{ i\sqrt{2\tka}\tG-\sqrt{2\tkd}\chi_a^{-1}[\omega] }{(\tOp - \omega)(\tOm - \omega)},\\\nonumber
    C^d_\R[\omega] &= \frac{i\sqrt{2\ga}\tG }{(\tOp - \omega)(\tOm - \omega)},
\end{align}
with the optical susceptibilities $\chi_a^{-1}[\omega] = \tka + \ga + i(\tDa - \omega)$ and $\chi_d^{-1}[\omega] = \tkd + i (\tDd - \omega)$.

The effective mechanical susceptibility $\chieff$ can be written
\begin{equation}\label{susceptibility}
    \chieff^{-1}[\omega] = \chi_{\m,0}^{-1}[\omega] + \chi_\opt^{-1}[\omega],
\end{equation}
where $\chi_{\m,0}[\omega] = \Om\left(\Om^2 - \omega^2 - i\omega\gam\right)^{-1}$ is the mechanical susceptibility of the bare resonator and $\chi_\opt[\omega]$ the optical contribution to the susceptibility, which reads
\begin{align}
    \chi_\text{opt}^{-1}[\omega]= -2\sum_{c} \left(\tgm^*C_q^c[\omega] +\tgm C_q^c[-\omega]^*\right),\label{Xopt}
\end{align}
with
\begin{align}\label{Cq_c}
    C^a_q[\omega] &= -i\frac{\chi_d^{-1}[\omega]\tga - i\tG \tgd}{(\tOp - \omega)(\tOm - \omega)},\\\nonumber
    C^d_q[\omega] &=-i\frac{\chi_a^{-1}[\omega]\tgd - i\tG \tga}{(\tOp - \omega)(\tOm - \omega)}.
\end{align}
These susceptibilities and their constituents deriving from different contributions of the dynamics, will play an important role in the characterization of the optomechanical response of the device in Sec.~\ref{sec:OM}.

\subsection{Studied devices}\label{sec:devices}

\begin{table*}
    \setlength{\tabcolsep}{3pt}
    \renewcommand{\arraystretch}{1.2}
    \begin{tabularx}{\linewidth}{llXXX}
    Parameter & Description &  Device (i) & Device (ii) & Device (iii) \\
    \hline\hline
    \multicolumn{5}{l}{\itshape Optical modes} \\ \hline
    $\Ddiff/2\pi$  & effective detuning between the cavity and mirror modes (\si{\hertz})                 & $-1.42\times 10^{10}$           & $\phantom{-}0$                  & $\phantom{-}1.97\times 10^{13}$ \\
    $\ga/2\pi$     & coupling between the cavity mode and the right environment (\si{\hertz})             & $\phantom{-}3.66\times 10^{11}$ & $\phantom{-}1.30 \times 10^{5}$ & $\phantom{-}6.00\times 10^{8}$   \\
    $\tka/2\pi$    & coupling between the cavity mode and the left environment (\si{\hertz})              & $\phantom{-}2.12\times 10^{12}$ & $\phantom{-}2.12\times 10^{12}$ & $\phantom{-}1.00\times 10^{13}$   \\
    $\tkd/2\pi$    & coupling between the mirror mode and the left environment  (\si{\hertz})             & $\phantom{-}3.80\times 10^{12}$ & $\phantom{-}2.12\times 10^{12}$ & $\phantom{-}1.08\times 10^{9}$  \\
    $\lambda/2\pi$ & coupling between the cavity and mirror modes (\si{\hertz})                           & $\phantom{-}4.09\times 10^{12}$ & $\phantom{-}4.09\times 10^{12}$ & $\phantom{-}4.09\times 10^{11}$   \\
    $\Plas$        & laser power used in Figs.~\ref{fig:g_eff}-\ref{fig:weak coupling} (\si{\micro\watt}) & $\phantom{-}150$                & $\phantom{-}50$                 & $\phantom{-}0.256$              \\
    \hline
    \multicolumn{5}{l}{\itshape Mechanical mode} \\ \hline
    $\Om/2\pi$     & bare mechanical frequency (\si{\hertz})                                              & $\phantom{-}5.14\times 10^5$     & $\phantom{-}1.3\times 10^6$     & $\phantom{-}1.3\times 10^6$     \\
    $\gam/2\pi$    & bare mechanical damping rate (\si{\hertz})                                           & $\phantom{-}17.1$               & $\phantom{-}9.3\times 10^{-3}$  & $\phantom{-}9.3\times 10^{-3}$  \\
    $Q_\m$         & mechanical quality factor                                                            & $\phantom{-}3\times 10^4$       & $\phantom{-}1.4\times 10^8$     & $\phantom{-}1.4\times 10^8$     \\
    \hline
    \multicolumn{5}{l}{\itshape Single-photon optomechanical couplings} \\ \hline
    $\gwa/\Om$     & cavity dispersive optomechanical coupling                                            & $\phantom{-}1.6$                & $\phantom{-}6.5\times 10^{-5}$  & $\phantom{-}6.5\times 10^{-5}$  \\
    $\gwd/\Om$     & mirror dispersive optomechanical coupling                                            & $-3.5$                          & $-1.4\times 10^{-4}$            & $-1.4\times 10^{-4}$            \\
    $\gka/\Om$     & cavity dissipative optomechanical coupling                                           & $\phantom{-}1.5$                & $\phantom{-}6.0\times 10^{-5}$  & $\phantom{-}6.0\times 10^{-5}$  \\
    $\gkd/\Om$     & mirror dissipative optomechanical coupling                                           & $\phantom{-}6.3$                & $\phantom{-}2.5\times 10^{-4}$  & $\phantom{-}7.1\times 10^{-8}$  \\
    \hline
    \multicolumn{5}{l}{\itshape Effective optical ``$-$'' mode } \\ \hline
    $\tkm/\Om$     & sideband resolution                                                                  & $\phantom{-}5.4\times 10^{5}$   & $\phantom{-}0.05$               & $\phantom{-}0.05$               \\
    $\gw[-]/\Om$   & single-photon dispersive ultrastrong coupling ratio                                  & $-0.97$                         & $-3.8\times 10^{-5}$            & $-1.4\times 10^{-4}$            \\
    $\gk[-]/\Om$   & single-photon dissipative ultrastrong coupling ratio                                 & $\phantom{-}0.065$              & $\phantom{-}2.9\times 10^{-12}$ & $\phantom{-}3.8\times 10^{-12}$ \\
    $\gw[-]/\tkm$  & single-photon dispersive strong coupling ratio                                       & $-1.8\times 10^{-6}$            & $-7.5\times 10^{-4}$            & $-2.8\times 10^{-3}$            \\
    $\gk[-]/\tkm$  & single-photon dissipative strong coupling ratio                                      & $-1.2\times 10^{-7}$            & $\phantom{-}5.7\times 10^{-11}$ & $\phantom{-}7.7\times 10^{-11}$ \\
    \hline\hline
    \end{tabularx}
    \caption{\label{tab:params}
        Parameters used for the three studied devices: (i) experimental device from Ref.~\cite{Manjeshwar2023May}, (ii) device with matching optical modes,  $\Ddiff = \kdiff = 0$ and (iii) device with  $\Ddiff = \kdiff \lambda/\sqrt{\tka\tkd}$. For Devices (ii) and (iii), we have chosen larger state-of-the art mechanical frequencies and quality factors \cite{Saarinen2023Mar}. The optical loss rates given here are the ones evaluated at the average mechanical displacement $\bq$ ($\tka, \tkd$) but for all three devices the difference with the parameters evaluated at zero displacement ($\ka, \kd$) is negligible.
        The laser powers indicated for Devices (ii) and (iii) are chosen such that the devices are in the weak-coupling regime to show the blue and red sidebands in Figs.~\ref{fig:dOm and Gopt} and \ref{fig:weak coupling}.
        Laser powers required for achieving ground-state cooling, as discussed in Sec.~\ref{sec:cooling:ground-state}, are however larger, of the order of $10$ \si{\watt} and $10$ \si{\milli\watt} respectively (see Fig.~\ref{fig:cooling}) in Sec.~\ref{sec:nonlinear}.
    }
\end{table*}

To give concrete examples and illustrate some of the properties of this complex optomechanical setup, we choose three sets of parameters, given in Table~\ref{tab:params}: (i) the experimental device from Ref.~\cite{Manjeshwar2023May}
(see reference for details about obtaining the parameters and confirming the optomechanical coupling strengths by other methods), (ii) a device with identical optical modes, i.e. $\Ddiff = \kdiff = 0$ and (iii) a device\footnote{Ref.~\cite{Cernotik2019Jun} studies a different case. First, there are no optomechanical effects on the Fano mode and no dissipative optomechanical effects on the cavity mode in ~\cite{Cernotik2019Jun}. Second, the cavity and mirror modes in~\cite{Cernotik2019Jun} are detuned by $\Ddiff = \lambda\sqrt{\ka/\kd}$, which comes from an assumption the authors made in their transfer matrix modeling, amounting to enforcing $\tDd \simeq \tDm$ and given their parameters (especially the free spectral range), they focus on longer cavities and their $\kd$ is a lot closer to the mechanical frequency compared to the values we consider in the following. While this choice of parameters is not specifically addressed in the present work, it can however be covered by our model.
} with $\Ddiff = \kdiff \lambda/\sqrt{\tka\tkd}$.

All three devices are in the strong optical coupling regime (see Fig.~\ref{fig:regimes} for a summary of the possible parameter regimes) and the effective ``$-$'' mode hence characterizes the physics of all three devices, while the highly damped ``$+$'' mode can be disregarded in the interpretation of the results. More precisely, Devices (i) and (ii) are in the regime $\abs{\tG} > \abs{\kdiff}, \abs{\Ddiff}, \tka, \tkd$ and, therefore, the effective modes ``$-$'' and ``$+$'' are very different from modes $a$ and $d$, both in term of frequency and loss rate.  On the other hand, Device (iii) is less strongly coupled in the sense that $\abs{\kdiff}, \abs{\Ddiff}, \tka > \abs{\tG} > \tkd$, such that only the loss rate $\tkm$ is strongly modified.
The parameters from (ii) and (iii) are inspired from (i) but modified to reach the effective resolved-sideband regime, $\tkm \ll \Om$, see also the discussion in the end of Sec.~\ref{sec:model:mech:micro}, and such that $\tkm^{(ii)} = \tkm^{(iii)}$.
Furthermore, to make it easier to reach the effective resolved-sideband regime and to enable ground-state cooling at room temperature, we choose state-of-the-art mechanical parameters, $Q_\m = 1.4\times 10^8$ and $\Om/2\pi = 1.3 $ \si{\mega\hertz}, from a SiN 2D phononic crystal membrane \cite{Saarinen2023Mar}. But similar numbers could be achieved with InGaP-based mechanics \cite{Manjeshwar2023Jun}, which is fully compatible with the microcavity from Ref.~\cite{Manjeshwar2023May}.

The frequency and loss rates of the mirror mode can be engineered by changing the photonic crystal pattern (lattice constant, hole radius) \cite{Fan2003Mar,Peralle2023Jun}, which makes it possible to realize the conditions $\Ddiff = \kdiff = 0$ and $\Ddiff = \kdiff \lambda/\sqrt{\tka\tkd}$ for Devices (ii) and (iii).
In addition, in order to reach the effective resolved-sideband regime, we need to identify appropriate damping rates $\gamma_a$ for the cavity-environment coupling via the right mirror. For Device (ii), we have $\tkm^{(ii)} \simeq \ga/2$ (see Sec.~\ref{sec:model:mech:Langevin}), meaning that we need to take $\ga \ll \Om$. While this might be challenging to achieve in specific microcavity setups of the type of Ref.~\cite{Manjeshwar2023May} (where in practice it would need to include not only the transmission through the right mirror, but also all other losses like absorption or scattering), Device (ii) still constitutes an intriguing alternative setting to Device (iii).
For Device (iii), we have the weaker constraint $\tkm^{(iii)} \simeq \ga\tkd/2\ktot$ (see Sec.~\ref{sec:model:mech:Langevin}), so by taking $\tka \gg \tkd$ we obtain a device reaching the resolved-sideband regime with parameters for $\gamma_a$ that are realistic for experimental realizations as in Ref.~\cite{Manjeshwar2023May}. Indeed, with a highly reflective right mirror, such as a distributed Bragg reflector, one can achieve a very low transmission of around 10 ppm and ideally get total losses (transmission through the right mirror, absorption and scattering) of 20 ppm. This would give a realistic value of $\ga/2\pi \simeq 600$ \si{\mega\hertz} for Device (iii).

The optomechanical couplings in Devices (ii) and (iii) were decreased by 4 orders of magnitude compared to Device (i). While this might deviate from expected values for currently realized microcavities, it ensures that those effectively sideband-resolved devices are in the linear regime (Sec.~\ref{sec:linearized om}) and hence fulfill the stability criteria of Appendix~\ref{app:stability}. In Device (iii), $\gkd$ was further decreased to match the decrease in $\tkd$ compared to Device (ii). In Sec.~\ref{sec:nonlinear}, we show how larger - and possibly even more realistic values for the optomechanical couplings - can allow for reaching the effective strong and ultrastrong coupling regimes.

Finally, in the following, we call ``standard device'' (denoted with superscript $\std$) a canonical optomechanical setup with identical frequency-independent mirrors and the same sideband resolution as Devices (ii) and (iii), namely $\ka^\std = \ga^\std = \tkm^{(ii)}/2$, no mirror mode $d$ and only a dispersive optomechanical coupling $\gwa$.

\section{Optomechanical properties}\label{sec:OM}

In this section, we study the optical and mechanical properties of the setup that can be measured in experiments so as to find signatures of the different optomechanical parameter regimes. We provide expressions to fit experimental data and determine the parameters characterizing the device.

\subsection{Effective optomechanical couplings}\label{sec:OM:geffs}

The values of the effective optomechanical couplings, Eq.~\eqref{effective couplings}, vary a lot depending on the system's parameters.
First, they are all proportional to $\aL$ and therefore increase like the square root of the input laser power $\Plas$.
For purely dispersive optomechanics, namely $\gka = \gkd = 0$, one gets $\tg = \tgm = \gw \bc$. In this case, like in standard optomechanics, each coupling strength is simply enhanced by the square root of the average photon number in the corresponding mode ${c} = {a}, {d}$.
However, the picture is more complicated for purely dissipative coupling. Even in the simple case $\gwa = \gwd = \gkd = 0$, $\tga$ and $\tgma$ are different while $\tgd = - \tgmd \neq 0$ since $\gadm = \gadp = \frac{1}{2}\sqrt{\frac{\kd}{\ka}}\gka$.

\begin{figure}
    \includegraphics[width=\linewidth]{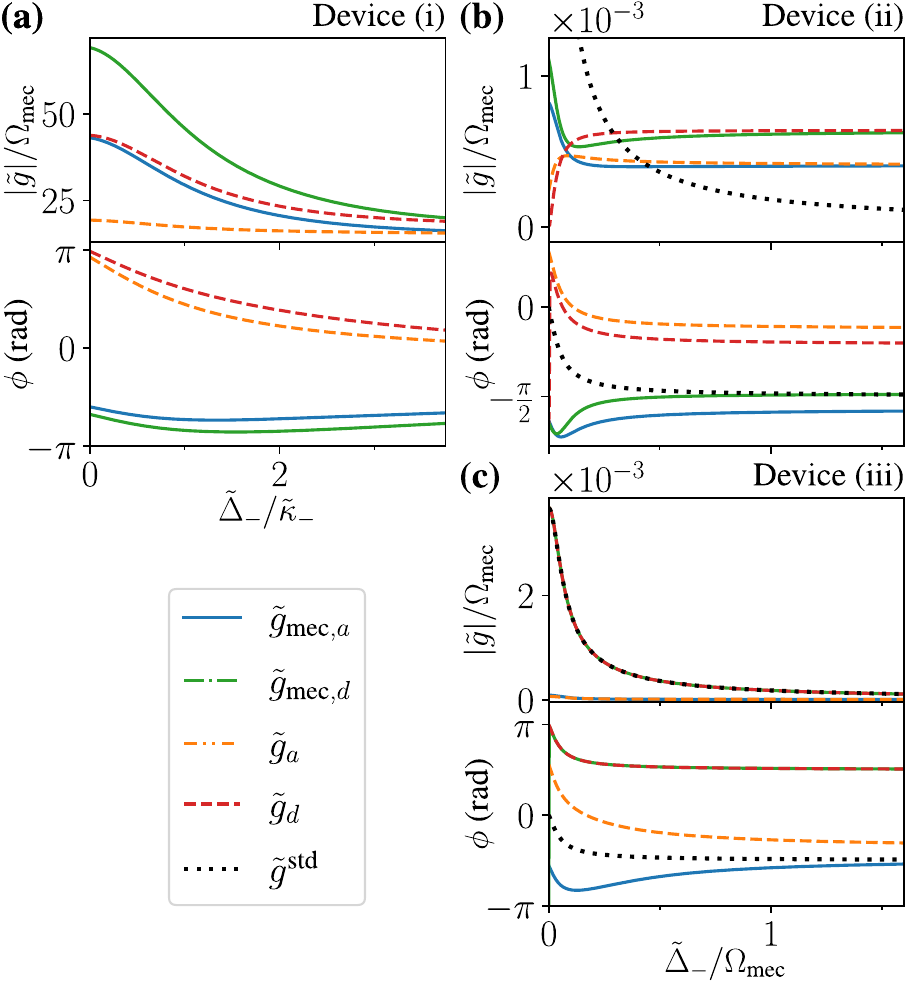}
    \caption{\label{fig:g_eff}
        Effective couplings (Eq.~\eqref{effective couplings}) as functions of detuning for \textbf{(a)} Device (i), \textbf{(b)} Device (ii), and \textbf{(c)} Device (iii). The couplings are complex numbers and the top panels show the absolute values of the couplings while the bottom panels show the phase, $\tg[] = \abs{\tg[]}\e^{i\phi}$. The dotted black line corresponds to the effective coupling $\tg[]^\std = \gwa \ba$ of a standard optomechanical device with the same sideband resolution as Devices (ii) and (iii).
        The parameters are given in Table~\ref{tab:params}.
    }
\end{figure}

In Fig.~\ref{fig:g_eff}, we show how the (complex-valued) couplings change with the \mbox{``$-$''-mode} detuning $\tDm$ for each device.
The change in their phases directly originates from the change of phase in $\ba$ and $\bd$ due to the modification of $\tDa$ and $\tDd$, see Eq.~\eqref{ss}. We see that Device (i) is in the ultrastrong coupling regime $\abs{\tilde{g}} > \Om$ as determined in Ref.~\cite{Manjeshwar2023Jun}, while Devices (ii) and (iii) are not, which is due to our choice of single-photon optomechanical coupling strengths (see Sec.~\ref{sec:devices} and Table~\ref{tab:params}).
For a standard optomechanical device with only dispersive optomechanical coupling and without mirror mode $d$ (dotted black line in Figs.~\subfigref{fig:g_eff}{b} and \subfigref{fig:g_eff}{c}), the optomechanical coupling is $\tga = \tgma = \gwa \ba \equiv \tg[]^\std$. Since, in this case, there is a single effective coupling, the phase of $\tg[]^\std$ does not matter and $\tg[]^\std$ can always be made real by changing the laser phase (in $\aL$).
This is clearly not the case for Device (ii) as there are four distinct effective couplings and some relative phases will always remain.
Conversely, for Device (iii), the effective optomechanical coupling of the mirror mode dominates, and since $\tgd \simeq \tgmd$, it behaves effectively like a standard device: the curves for $\abs{\tgd}$, $\abs{\tgmd}$ and $\abs{\tg[]^\std}$ are superimposed in Fig.~\subfigref{fig:g_eff}{c}.

\subsection{Mean optical response}\label{sec:OM:optical_response}

The mean steady-state optical output of the system can be measured in an experiment, especially the intensity transmission and reflection coefficients, $T(\omL) = \abs{t(\omL)}^2$ and $R(\omL) = \abs{r(\omL)}^2$. Here, we have denoted  $t = \mean{\boutR}/\mean{\binL}$ the amplitude transmission coefficient, from left to right since we have $\mean{\binR} = 0$, and the amplitude reflection coefficient, $r = \mean{\boutL}/\mean{\binL}$. Using the input-output relation \eqref{input-output}, we obtain
\begin{align}
    t &= -\sqrt{2\ga}\frac{\ba}{\aL} = -\sqrt{2\ga}C^a_\L[0],\\\nonumber
    r &= 1-\sum_c\sqrt{2\tk}\frac{\bc}{\aL} = 1-\sum_c\sqrt{2\tk}C^c_\L[0],\\\nonumber
\end{align}
with the steady-state amplitudes given in Eq.~\eqref{ss} and the coefficient $C^c_\L$ defined in Eqs.~\eqref{C_mu}. With this, the intensity transmission becomes
\begin{equation}
    T = \frac{2\ga}{D}\left(\sqrt{2\tka}\tDd - \sqrt{2\tkd}\lambda\right)^2,
\end{equation}
with
\begin{align}
    D =\, & \abs{\tOm}^2\abs{\tilde\Omega_+}^2 = \left(\tDm^2 + \tkm^2\right) \left(\tDp^2 + \tk[+]^2\right)\ .
\end{align}
The intensity transmission is modulated by the average mechanical displacement in the following way,
\begin{equation}\label{dTdq}
   \diff{T}{\bq} =\sqrt{2} \sum_c \left(-\gw \diffp{T}{\tDc} + \gk\diffp{T}{\tk}\right),
\end{equation}
see Appendix \ref{app:mean opt} for the full expressions of the derivatives.

\begin{figure}
    \includegraphics[width=\linewidth]{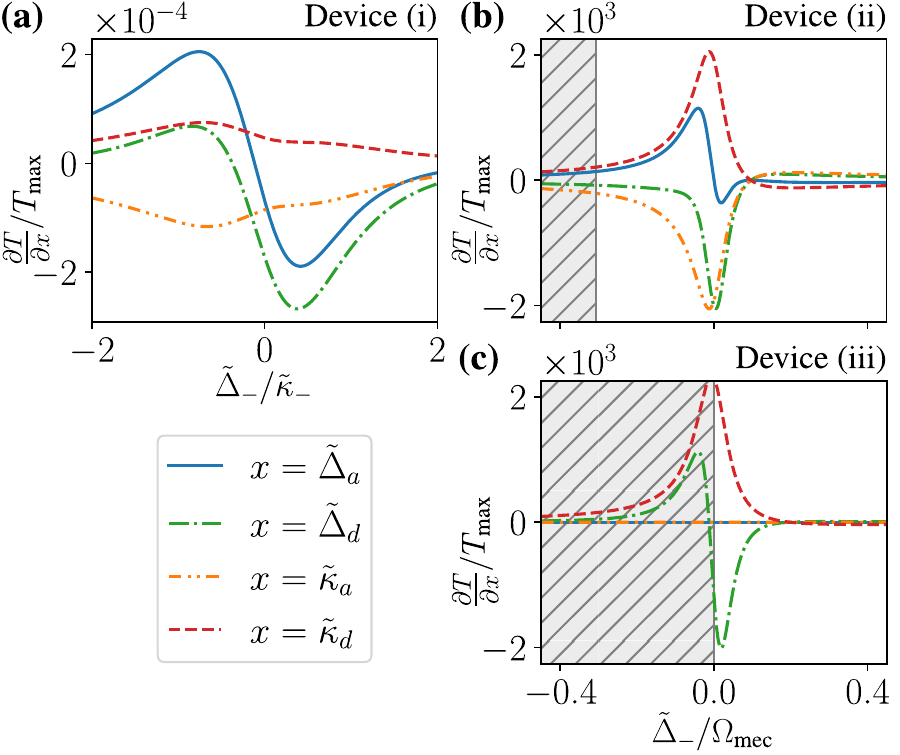}
    \caption{\label{fig:dTs}
        Optical response to the variation of different detunings and bath couplings as a function of the ``$-$''-mode detuning
        for \textbf{(a)} Device (i), \textbf{(b)} Device (ii) and \textbf{(c)} Device (iii). We have plotted each partial derivative from Eq.~\eqref{dTdq}. The hatched areas indicate detunings for which the respective devices are unstable due to heating (see Fig.~\subfigref{fig:dOm and Gopt}{a} and Appendix \ref{app:stability}). The parameters are given in Table~\ref{tab:params}.
    }
\end{figure}

In Fig.~\ref{fig:dTs}, we plot the constituents of Eq.~\eqref{dTdq}, namely the partial derivatives associated with each kind of optomechanical coupling.
We first notice that the main response takes place around $\tDm = 0$, that is when the laser is close to resonance with the ``$-$'' effective optical mode.
We have normalized $\difsp{T}{x}$, with $x = \tDa, \tDd, \tka, \tkd$, by the maximum transmission of the optical setup $T_\text{max}$. The motivation for this is that the devices have very different $\ga$ and, in particular, Device (ii) barely transmits any light since the coupling to the right environment is many orders of magnitude smaller than the couplings to the left environment, see Table~\ref{tab:params}. Fig.~\ref{fig:dTs} shows that the transmission of Device (i) is affected very little by the mechanics since it is very far from the effective sideband-resolved regime, unlike for the other devices.
We also see that, for Devices (i) and (ii), all the derivatives have similar magnitudes though different shapes, while, for Device (iii) in Fig.~\subfigref{fig:dTs}{c}, the derivatives with respect to the Fano-mirror-mode quantities largely dominate, as already observed for the effective optomechanical couplings.
For a standard optomechanical device, the only relevant derivative is $\difsp{T^\std}{\tDa}$, which is similar to $\difsp{T}{\tDd}$ for Device (iii) (dashed-dotted green line in panel (c)), with two off-resonant sidebands \cite{Baraillon2020Sep}. The main difference between the two is that $\difsp{T^\std}{\tDa}$ is an odd function of the detuning $\tDa$, while $\difsp{T}{\tDd}$ is not exactly an odd function of $\tDm$ due to the coupling between the optical modes. Similarly, there is an asymmetry in the dominant dissipative derivative $\difsp{T}{\tkd}$ while dissipative derivatives would be even functions of the detuning in the absence of coupling between the optical modes \cite{Baraillon2020Sep}.

\subsection{Mechanical response}\label{sec:OM:mechanical_response}

After studying how the mechanical motion impacts the mean optical transmission, we now do the opposite and analyze the effects of the optomechanical couplings on the mechanical motion.
To do so, we write the effective mechanical susceptibility, Eq.~\eqref{susceptibility}, in the usual form for a harmonic oscillator~\cite{Aspelmeyer2014Dec}, that is $ \chieff[\omega] = \Om\left(\Om^\eff[\omega]^2 - \omega^2 - i\omega\gamma^\eff[\omega]\right)^{-1}$, identifying the effective mechanical frequency $\Om^\eff[\omega] = \Om + \dOm[\omega]$ and damping rate $\gamma^\eff[\omega] = \gam + \Gopt[\omega]$, where we have defined the optical contribution to the mechanical damping rate and the mechanical frequency shift, in the limit $\dOm \ll \Om$,
\begin{align}\label{dOm_Gopt}
    \Gopt[\omega]& = -\frac{\Om}{\omega} \Im(\chi_\opt^{-1}[\omega]),\\\nonumber
    \dOm[\omega]& = \frac{1}{2}\Re(\chi_\opt^{-1}[\omega]).
\end{align}

\begin{figure}
    \includegraphics[width=\linewidth]{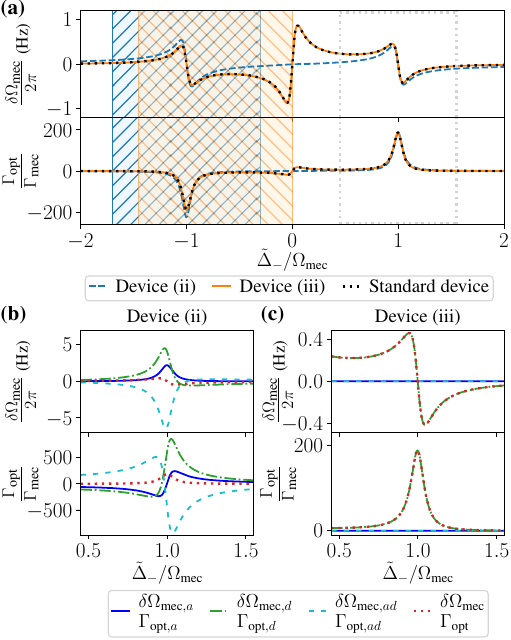}
    \caption{\label{fig:dOm and Gopt}
      \textbf{(a)} Mechanical frequency shift $\dOm$ (top panel) and optical contribution to the mechanical damping rate $\Gopt$ (bottom panel) as a function of the detuning $\tDm$ for Device (ii) and  Device (iii). The plots were obtained by evaluating Eqs.~\eqref{dOm_Gopt} at $\omega = \Om$.
       The hatched areas indicate detunings for which the respective devices are unstable due to the heating. The dotted black lines represent a standard resolved-sideband optomechanical device where the laser power has been adjusted to get a similar cooling on the red sideband.
    \textbf{(b)}, \textbf{(c)} Components of the mechanical frequency shift $\dOm$ (top panels) and optical contribution to the mechanical damping rate $\Gopt$ (bottom panels) as a function of the detuning $\tDm$ around the red sideband (gray dotted box in (a)) for  Device (ii) and Device (iii)  at $\omega = \Om$, see Eq.~\eqref{Xopt_comp}. The parameters are given in Table~\ref{tab:params}.
    }
\end{figure}

We have plotted $\dOm[\Om]$ and $\Gopt[\Om]$ for Device (ii) (in blue) and Device (iii) (in orange) as a function of the detuning $\tDm$ in Fig.~\subfigref{fig:dOm and Gopt}{a}. They give the frequency shift and optical contributions to the mechanical damping rate in the weak-optomechanical-coupling regime $\abs{\tg[]} < \Om, \tkm$, which is relevant for the laser powers considered here (see values in Table~\ref{tab:params} and Fig.~\ref{fig:g_eff} for $\abs{\tg[]}/\Om$).
We see frequency shifts on both sidebands, i.e. around $\tDm = \pm \Om$, which are similar to the ones of a standard optomechanical device, indicated by the dotted black line. Similarly, we observe the expected peaks in $\Gopt$ indicating respectively cooling on the red sideband and heating on the blue sideband.
For the considered laser powers, this heating effect is large enough to make the system unstable (hatched areas), see stability analysis in Appendix \ref{app:stability} for more details. This means that while the theory developed in this article, and in particular the linearization from Sec.~\ref{sec:linearized om}, is not valid in those hatched areas, it still predicts an overall negative mechanical damping rate\footnote{See, e.g., Ref.~\cite{Aspelmeyer2014Dec} for a discussion of this lasing effect, that we will not study in this work, for a standard optomechanical device.}.
All in all, both devices behave like a standard optomechanical device. However, while the behavior of Device (iii) is really identical to the one of a standard device,  Device (ii) has a different behavior around zero detuning.

We would like to note that Device (i) is not sideband-resolved, therefore the relevant detuning scale for the atypical frequency shift $\dOm[\Om]$ is $\tkm$ and not $\Om$ (discussed in detail in Ref.~\cite{Manjeshwar2023May}). The damping rate $\Gopt[\Om]$ of Device (i) is negligible, therefore there is no significant cooling or heating. These findings are reflected in the noise power spectrum of the mechanical position (Fig.~\subfigref{fig:PSD}{a}), discussed in the next subsection, Sec.~\ref{sec:OM:PSD}.

In order to better understand the contribution of the different optical modes to the mechanical damping rate and frequency shift, we now split the optical contribution to the mechanical susceptibility, using Eqs.~\eqref{Xopt} and \eqref{Cq_c}, into
\begin{equation}\label{Xopt_comp}
    \chi_{\opt}^{-1}[\omega] = \chi_{\opt,a}^{-1}[\omega]+\chi_{\opt,d}^{-1}[\omega]+\chi_{\opt,ad}^{-1}[\omega],
\end{equation}
with
\begin{align}
    \chi_{\opt,a}^{-1}[\omega] &=  -2i\frac{ \tgma^*\tga\chi_d^{-1}[\omega]}{(\tOp - \omega)(\tOm - \omega)} + (\omega\to -\omega)^*, \\\nonumber
    \chi_{\opt,d}^{-1}[\omega] &=  -2i\frac{ \tgmd^*\tgd\chi_a^{-1}[\omega]}{(\tOp - \omega)(\tOm - \omega)} + (\omega\to -\omega)^*, \\\nonumber
    \chi_{\opt,ad}^{-1}[\omega] &= -2\frac{   (\tgma^*\tgd + \tgmd^*\tga)\tG}{(\tOp - \omega)(\tOm - \omega)} + (\omega\to -\omega)^*.
\end{align}
The notation $\bullet + (\omega\to -\omega)^*$ means that we replace $\omega$ by $-\omega$ in $\bullet$ and take the complex conjugate. We can apply the same decomposition to the damping rate and the mechanical frequency shift, $\Gopt =  \Gamma_{\opt,a}+ \Gamma_{\opt,d}+ \Gamma_{\opt,ad}$ and $\dOm = \delta\Omega_{\m,a}+\delta\Omega_{\m,d}+\delta\Omega_{\m,ad}$, by applying the definitions \eqref{dOm_Gopt} to the components of $\chi_{\opt}^{-1}$.
From the effective coupling perspective, $\chi_{\opt,c}^{-1}[\omega]$ is a pure contribution from the optical mode ${c} = {a}, {d}$, but note that $\tg$ and $\tgm$ contain some cross-terms in $\gadm$ or $\gadp$, see Eqs.~\eqref{effective couplings}. However, in the limit of purely dispersive optomechanical couplings, namely $\gka, \gkd \to 0$, the factor $\tgm^*\tg$ becomes $(\gw\abs{\bc})^2$.

These three components of $\dOm[\Om]$ and $\Gopt[\Om]$ are plotted around the red sideband in Fig.~\subfigref{fig:dOm and Gopt}{b,c}. For Device (ii) in panel (b), the components have similar magnitudes and partially cancel each other out, giving rise to a lower total quantity (dotted red line). On the other hand, panel (c) confirms that the dominant contribution to the optomechanics in Device (iii) comes from the effective $d$ coupling since the dash-dotted green line representing the $d$ contribution is superimposed with the dotted red line of the total quantity.

\subsection{Quadrature output power spectra}\label{sec:OM:PSD}

In experiments, one typically measures the noise power spectrum of the light transmitted ($\aoutR$) or reflected ($\aoutL$) by the cavity, for instance with a homodyne detection scheme \cite{Manjeshwar2023May}, with the goal to also deduct information about the mechanics. Using the input-output relations Eq.~\eqref{input-output lin}, we compute the power spectra of the quadratures of the light leaking out of the cavity, defined by\footnote{Note that the spectra relevant for measurements such as homodyne detection are the symmetrized spectra. Due to the noise model considered in this work, we have $S_{Q_{\Out,\mu}}[-\omega]=S_{Q_{\Out,\mu}}[\omega]$ and the result of Eq.~\eqref{power spectrum1} hence is identical to the symmetrized version, see Appendix \ref{app:power spectra}. }
\begin{equation}
    S_{Q_{\Out,\mu}}[\omega]= \int_{-\infty}^{+\infty} \frac{d\omega'}{2\pi}\mean{\hat{Q}_{\Out,\mu}[\omega]\hat{Q}_{\Out,\mu}[\omega']},\label{power spectrum1}
\end{equation}
with $Q = X,P$ and $\mu=\L,\R$. Here, $\Xout = (\aout + \aout^\dagger)/\sqrt{2}$ and $\Pout = (\aout - \aout^\dagger)/i\sqrt{2}$ are the position and momentum quadratures of the output light.
We obtain
\begin{align}
    S_{X_{\Out,\L}}[\omega] =\,& S_{X_{\text{in},\L}}[\omega] + 2c_P^2S_q[\omega] + 4\sqrt{\tka\tkd}S_{X_a X_d}[\omega]\nonumber\\
    & + \sum_c \left(2\tk S_{X_c}[\omega] + 4\sqrt{\tk}c_P S_{X_c q}[\omega] \right) ,\nonumber\\
    S_{P_{\Out,\L}}[\omega] =\,& S_{P_{\text{in},\L}}[\omega] + 2c_X^2S_q[\omega] + 4\sqrt{\tka\tkd}S_{P_a P_d}[\omega]\nonumber\\
    & + \sum_c\left( 2\tk S_{P_c}[\omega]+  4\sqrt{\tk}c_X S_{P_c q}[\omega]\right) \nonumber\\
    S_{X_{\Out,\R}}[\omega] =\,& S_{X_{\text{in},\R}}[\omega] + 2\ga S_{X_a}[\omega] ,\nonumber\\
    S_{P_{\Out,\R}}[\omega] =\,& S_{P_{\text{in},\R}}[\omega] + 2\ga S_{P_a}[\omega], \label{output spectra}
\end{align}
with $S_Q[\omega]$ the spectrum of $\dQ$ for $Q = X_c, P_c, q$ and the input vacuum noise spectra $S_{X_{\In,\mu}}[\omega] = S_{P_{\In,\mu}}[\omega] = 1/2$.
We have also defined the cross terms
\begin{equation}
    S_{Q_1 Q_2}[\omega] = \frac{1}{2}\int_{-\infty}^{+\infty} \frac{\dl \omega'}{2\pi} \mean{\dQ_1[\omega]\dQ_2[\omega'] +\dQ_2[\omega]\dQ_1[\omega']}.
\end{equation}
These spectra can be computed from the solution of the Langevin equations \eqref{Langevin lin eqs}, see analytical expressions in Appendix \ref{app:power spectra}. In particular, for the mechanical position spectrum $S_q[\omega]$, using Eq.~\eqref{solution Langevin:q}, we get
\begin{equation}
    S_{q}[\omega] = \abs{\chi_\m^\eff[\omega]}^2(S_\text{th}[\omega] + S_\text{om}[\omega]), \label{Sq}
\end{equation}
where $S_\text{th}[\omega] \simeq 2\gam \Nm$ is the thermal noise spectrum in the high mechanical temperature limit ($\hbar\Om \ll \kB T_\m$) relevant here, see Ref.~\cite{Genes2008Mar} for a more general expression. Finally,
\begin{align}\label{Som}
    S_\text{om}[\omega] =\,&\frac{1}{2}\abs*{\sum_{c}2\tgm^* C^c_\L[\omega] + \sqrt{2} \left(c_X  -i c_P\right) }^2\\\nonumber
    & +\frac{1}{2}\abs*{\sum_{c}2\tgm^* C^c_\R[\omega]}^2
\end{align}
is the noise power spectrum due to the optomechanical effects.

\begin{figure}[tb]
    \includegraphics[width=\linewidth]{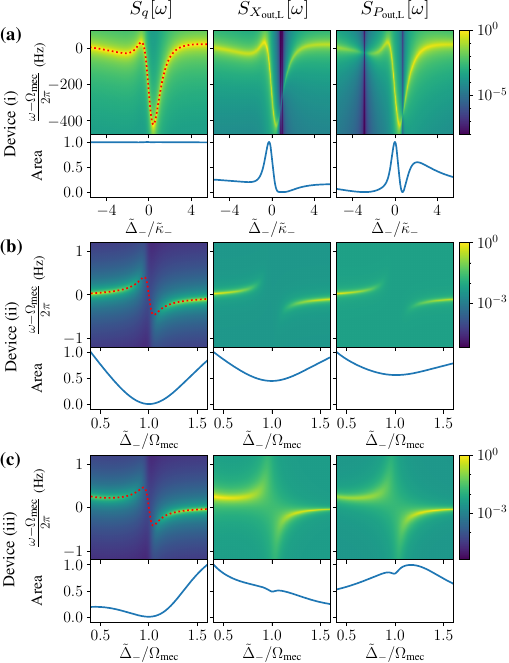}
    \caption{\label{fig:PSD}
        Noise power spectra of \textbf{(a)} Device (i) \textbf{(b)} Device (ii) and \textbf{(c)} Device (iii). \textit{Upper row of each panel:} Noise power spectrum of the mechanical position, $S_q[\omega]$, and light output quadratures on the left side, $ S_{X_{\Out,\L}}[\omega]$ and  $S_{P_{\Out,\L}}[\omega]$, as functions of the effective detuning $\tDm$ and of the frequency $\omega$. The dotted red line shows the mechanical frequency shift $\dOm[\Om]$. \textit{Lower row of each panel:} area corresponding to the integral over $\omega$ of the corresponding spectrum in the upper row.
        All the plotted quantities have been normalized by their maximum values; all parameters are given in Table~\ref{tab:params}.
    }
\end{figure}

In the high mechanical temperature limit, the part of the noise power spectra coming from the optical environments (vacuum noise) is negligible compared to the thermal noise from the mechanical environment.
As a consequence the spectrum given in Eq.~\eqref{Sq} can be approximated by $S_q[\omega] \simeq 2\gam\Nm \abs{\chi_\m^\eff[\omega]}^2$.
The noise spectra  $S_{X_{\Out,\mu}}[\omega]$ and $S_{P_{\Out,\mu}}[\omega]$ are proportional to the factor $\gam  \abs{\chi_\m^\eff[\omega]}^2$ as well and, therefore, exhibit a peak at the effective mechanical frequency, as shown in Fig.~\ref{fig:PSD} for all three devices (the effective mechanical frequency is indicated by the dotted red line).
However, the noise spectrum $S_{Q_{\Out,\mu}}[\omega]$, with $Q = X,P$, in general also has additional frequency-dependent factors and hence its area $\int \frac{\dl \omega}{2\pi} S_{Q_{\Out,\mu}}[\omega]$ does not give direct access to the mechanical fluctuations $\mean{\dq^2} = \int_{-\infty}^{+\infty} \frac{\dl \omega}{2\pi} S_q[\omega]$, see Appendix \ref{app:power spectra} for details on how to compute $\mean{\dq^2}$ from the optical output spectra.

For the devices we consider here, the frequency dependence of these factors can be neglected on the range of values of $\omega$ relevant for mechanical features, such that $\int \frac{\dl \omega}{2\pi} (S_{Q_{\Out,\mu}}[\omega] - S_{Q_{\In,\mu}}[\omega]) \propto \mean{\dq^2}$.
But this proportionality factor is strongly dependent on the detuning $\tDm$, which explains the difference between the area plots of optical and mechanical noise power spectra (lower row of each panel) in Fig.~\ref{fig:PSD}. In particular, for Device (i), these additional detuning-dependent contributions have an important effect that should not be confused with optomechanical cooling.
In Fig.~\subfigref{fig:PSD}{a}, both $S_{X_{\Out,\L}}[\omega]$ and $S_{P_{\Out,\L}}[\omega]$ exhibit dark blue lines at distinct values of $\tDm$, where the optical spectra are suppressed. This is in qualitative agreement with the experimental measurements \cite{Manjeshwar2023May}.
Note that this is a pure optical effect, as can be seen from the area of the mechanical spectrum $S_q[\omega]$, $\int \frac{\dl \omega}{2\pi} S_q[\omega]$, in the lower-left panel of Fig.~\subfigref{fig:PSD}{a}, which is constant. Conversely, for the other two devices, the cooling observed as a dip of the area of $S_q[\omega]$ around $\tDm = \Om$ is still visible on top of the optical contributions to the areas of $S_{X_{\Out,\mu}}[\omega]$ and $S_{P_{\Out,\mu}}[\omega]$. In Fig.~\subfigref{fig:PSD}{b}, for Device (ii), the optical contributions manifest as an approximately constant shift of the area while for Device (iii), in Fig.~\subfigref{fig:PSD}{c}, the cooling dip is visible---even though only faintly---on top of the $\tDm$-dependent optical features.

\section{Back-action cooling}\label{sec:cooling}

After having demonstrated the general properties of the optomechanical Fano-mirror system, we now focus on a widely used application, namely cooling of the mechanical resonator using optomechanical back-action. It is well known that this cooling scheme allows to bring \emph{sideband-resolved} standard optomechanical systems into the mechanical ground state \cite{Teufel2011Jul, Chan2011Oct, Delic2020}. In this section, we show that similar achievements are possible for our system, even if $\kd, \ka + \ga \gg \Om$ [Device (ii)] or  $\kd, \ka, \ga \gg \Om$ [Device (iii)].
This is possible since these devices are in the \emph{effective} sideband-resolved regime $\tkm \lesssim \Om$.

We start this section by analyzing the underlying cooling mechanisms in the weak-coupling regime, where the optomechanical coupling strengths are significantly smaller than the mechanical frequency and the (effective) optical loss rates, which is the case for the laser powers given in Table~\ref{tab:params}.

\subsection{Stokes and anti-Stokes scattering processes}\label{sec:cooling:weak coupling}

The steady-state phonon number in the mechanical fluctuations,
\begin{equation}
    \Neff = \frac{1}{2}(\mean{\dq^2} + \mean{\deltp^2} - 1),\label{Neff}
\end{equation}
results, in the weak optomechanical-coupling limit, from the competition between the Stokes and anti-Stokes scattering processes, which respectively create and annihilate phonons in the mechanical resonator.
The respective rates $A_+$ and $A_-$ of those processes are given by $A_\pm = \frac{1}{2}S_{FF}[\mp\Om]$ \cite{Aspelmeyer2014Dec}, where $S_{FF}[\omega] =  \int_{-\infty}^{+\infty} \frac{d\omega'}{2\pi}\mean{\delta\hat{F}[\omega]\delta\hat{F}[\omega']}$ is the noise power spectrum of the optical back-action on the mechanical resonator,
\begin{align}
    \delta\hat{F} =\,& \sum_{c}\sqrt{2} (\tgm \dc^\dagger +\tgm^*\dc)\\\nonumber&+ c_X\sqrt{2}\XinL + c_P\sqrt{2}\PinL,
\end{align}
obtained from the steady state-solution of the Langevin equations, see the last two lines of Eq.~\eqref{Langevin lin eqs}.
In the weak-coupling limit, the effect of the mechanics on the optics can be neglected and we compute $S_{FF}[\omega]$ by solving the Langevin equations~\eqref{Langevin lin eqs} for the optics only, see also Eq.~\eqref{Langevin optics}. This yields $\dc[\omega] = C_\L^c[\omega]\ainL[\omega] +  C_\R^c[\omega]\ainR[\omega] $ and hence,
\begin{align}
    A_\pm =\,& \abs*{\sum_c \left(\tgm^* C_\L^c[\mp\Om] - \frac{\gk\aL^*}{\sqrt{2\kc}}C_\L^c[0]\right)}^2\\\nonumber
    &+\abs*{\sum_c \tgm^* C_\R^c[\mp\Om] }^2.
\end{align}
From these rates, we identify the characteristics of the \emph{effective phonon bath} created by the optical part of the setup \cite{Monsel2021Jun, Aspelmeyer2014Dec}: its effective damping rate $\Gopt^\wc= A_- - A_+$ (coinciding with $\Gopt[\Om]$ from Fig.~\subfigref{fig:dOm and Gopt}{a} in this limit) and its effective average phonon number $\Nopt = A_+/\Gopt$. As a result of the coupling to this effective phonon bath created by the optical setup, the steady-state phonon number in the mechanical fluctuations is then given by
\begin{equation}\label{Neff_wc}
   \Neff^\wc=\frac{\gam\Nm + \Gopt^\wc \Nopt}{\gam + \Gopt^\wc}.
\end{equation}

\begin{figure}
    \includegraphics[width=\linewidth]{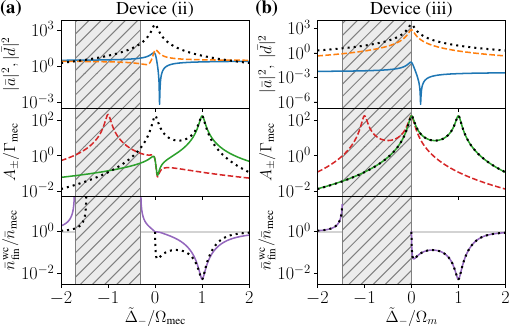}
    \caption{\label{fig:weak coupling}
        Optomechanical cooling in the weak-coupling limit and resolved sideband regime $\tkm \ll \Om$. We show results for \textbf{(a)} Device (ii) and \textbf{(b)} Device (iii), see parameters in Table~\ref{tab:params}.
        Top panels: average photon numbers in the cavity ($\abs{\ba}^2$, solid blue) and in the Fano mirror ($\abs{\bd}^2$, dashed orange); center panels: anti-Stokes (solid green) and Stokes (dashed red) rates; bottom panels: obtained average phonon number in the mechanical fluctuations (solid purple), the gray line indicates the thermal phonon number $\Nm$. All functions are plotted in dependence of the detuning between the laser and the effective ``$-$'' mode.
        The dotted black lines correspond to the result for a standard optomechanical device with the same sideband resolution (we have only plotted $A_-$ in the middle panel), and the hatched areas indicate detunings for which the system is unstable due to heating.
    }
\end{figure}

In the middle panels of Fig.~\ref{fig:weak coupling}, we see that the Stokes rate $A_+$ (dashed red) and anti-Stokes rate $A_-$ (solid green) exhibit peaks at the blue and red sidebands respectively, corresponding to heating and cooling.
Device (ii) (Fig.~\subfigref{fig:weak coupling}{a}) has similar features as a standard resolved-sideband optomechanical device but with a slight asymmetry between $A_+$ and $A_-$, which explains why $\Gopt$ is not an odd function of the detuning in Fig.~\subfigref{fig:dOm and Gopt}{a}.
For both devices, the weak optomechanical-coupling approximation gives a good estimate of $\Neff$
(see purple diamonds in Fig.~\ref{fig:cooling}(a) and (b)). Figure~\ref{fig:weak coupling} confirms the observation from Fig.~\subfigref{fig:dOm and Gopt}{a} that, despite the large parameter differences between Device (ii) and Device (iii), their behavior on both sidebands are very similar, almost identical to the one of a standard device (dotted black line). Differences mainly occur close to zero detuning, which is not relevant for cooling applications.

The effective optomechanical couplings, $\tgm$ are proportional to the laser drive amplitude $\aL$, see Eqs.~\eqref{ss} and Eqs.~\eqref{effective couplings}, such that $A_\pm$ and $\Gopt^\wc$ are proportional to the laser power as well. Consequently, increasing $\Plas$ brings the steady-state phonon number $\Neff^\wc$, Eq.~\eqref{Neff_wc}, closer to $\Nopt$. Since $\Nopt \simeq 6.3\times 10^{-4}\ll 1$ for Devices (ii) and (iii), ground-state cooling is expected to be possible. However, the weak-coupling approximation typically breaks down well before $\Neff$ reaches $\Nopt$. We therefore need to study the steady-state phonon number beyond the weak-coupling limit, as will be done in the following subsection~\ref{sec:cooling:ground-state}.

\subsection{Ground-state cooling}\label{sec:cooling:ground-state}

In the limit where the weak optomechanical-coupling approximation fails, the phonon number in the mechanical fluctuations, $\Neff$, can be computed from Eq.~\eqref{Neff} by solving the Lyapunov equation for the system in the steady state, as detailed in Appendix \ref{app:Lyapunov}.

\begin{figure}[t]
    \includegraphics[width=\linewidth]{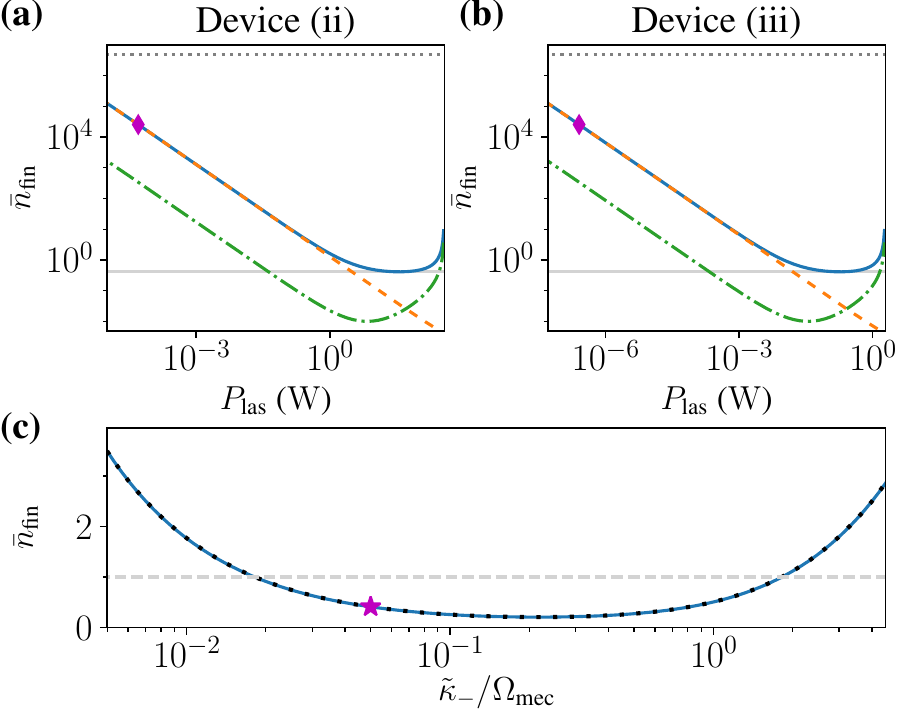}
    \caption{\label{fig:cooling}
        Steady-state phonon number in the mechanical fluctuations (Eq.~\eqref{Neff}).  Results as function of the laser power at $\tDm = \Om$ are shown for \textbf{(a)} Device (ii)
        and \textbf{(b)} Device (iii) at room temperature, $T_\m = 300$ K (solid blue line), and at cryogenic temperature, $T_\m=4$ K (dash-dotted green line). For the room-temperature case, the dotted gray line indicates the thermal phonon number, the dashed orange line the weak-coupling approximation (Eq.~\eqref{Neff_wc}) and the solid gray line the minimum reachable phonon number. The purple diamonds indicate the laser powers used in the other figures as indicated in Table~\ref{tab:params}. \textbf{(c)} Minimum steady-state phonon number in the mechanical fluctuations (minimized on $\tDm$ and $\Plas$) as function of the sideband resolution, where $\tkm$ is tuned via $\tkd$ and $\tDd$. The solid blue line was obtained with a device differing from Device (iii) only by the laser power with $\Ddiff = \kdiff \lambda/\sqrt{\tka\tkd}$. The purple star indicates Devices (ii) and (iii) and the dotted black line corresponds to the minimum $\Neff$ for a standard optomechanical device, with the ideal parameters $\ga = \ka = \tkm/2$. The dashed-gray line indicated the ground-state cooling threshold. All other parameters are given in Table~\ref{tab:params}.}
\end{figure}

We plot $\Neff$ obtained with optomechanical cooling as a function of the laser power for Devices (ii) and (iii) in Fig.~\subfigref{fig:cooling}{a,b}, where we choose the detuning $\tDm = \Om$ due to the weak-coupling results from Sec.~\ref{sec:cooling:weak coupling}.
We see that both Fano-mirror devices (ii) and (iii) can achieve ground-state cooling, both for cryogenic temperature, at $T_\m = 4$ K (dash-dotted green lines) and for room temperature (solid blue lines). The phonon number reaches a minimum of $\Neff \simeq 0.41$ at room temperature, while a lower phonon occupation, $\Neff \simeq 0.01$ can be reached at cryogenic temperature. In Appendix~\ref{app:GS cooling}, we additionally show that energy equipartition, $\mean{\dq^2} \simeq \mean{\deltp^2}$, is satisfied close to the lowest phonon occupation. As shown by the dashed orange line for the room-temperature case, the weak-coupling approximation eventually breaks down at large laser powers.

Even though Devices (ii) and (iii) have very different optical parameters for modes $a$ and $d$, panels (a) and (b) in Fig.~\ref{fig:cooling} are almost identical, showing that the key optical parameter determining the cooling limit is the effective sideband resolution $\tkm/\Om$. Therefore, we have also plotted the minimum phonon number as a function of the sideband resolution in Fig.~\subfigref{fig:cooling}{c}, showing that devices differing from Device (iii) in Table~\ref{tab:params} only by the tuneable laser power (solid blue line) have identical cooling performances as a standard optomechanical device (dotted black line). This further confirms the results from Figs.~\ref{fig:dOm and Gopt} and \ref{fig:weak coupling}, namely that around the red sideband, this complex engineered device behaves like a simple standard resolved-sideband optomechanical device, where the ideal sideband resolution has been achieved thanks to the Fano-mirror engineering.

Evidently, the mechanical frequency and quality factor also play a major role in determining the cooling limit. Room-temperature ground-state cooling is made possible here by the large mechanical frequency and high mechanical quality factor (see Table~\ref{tab:params}). In addition, note that we have assumed the drive laser to be shot noise limited, that is in a coherent state, while in an experiment, one would need to account for the laser phase noise and make sure it does not prevent the setup from reaching the mechanical ground state \cite{Rabl2009May, Kippenberg2013Jan}. Finally, the only difference between panels (a) and (b) in Fig.~\ref{fig:cooling} is the laser power at which the minimum $\Neff$ is reached, Device (ii) requiring a laser power two orders of magnitude larger than Device (iii). This comes from the difference in the optical parameters, in particular giving rise to different effective optomechanical couplings $\gw[-]$ and different laser frequencies giving $\tDm = \Om$. Therefore identical laser powers correspond to different input photon rates $\Plas/\hbar\omL$.

\section{Towards nonlinear optomechanics}\label{sec:nonlinear}

In this section we show that a system, as realized by a microcavity with a frequency-dependent Fano mirror, can reach the effective resolved-sideband, strong and ultrastrong coupling regimes simultaneously.
To that end, we introduce two additional model systems, Devices (iv) and (v), with parameters that are close to situations that can be experimentally achieved with microcavities~\cite{Manjeshwar2023Jun,Peralle2023Jun}, see Table~\ref{tab:params2}.
These systems are based on state-of-the-art mechanical parameters \cite{Saarinen2023Mar} and, like Device (iii), fulfill the condition $\Ddiff = \kdiff \lambda/\sqrt{\tka\tkd}$. It allows them to be in the effective resolved-sideband regime with $\tkm < \Om$ obtained from Eq.~\eqref{eq:eigenfreq}.

\begin{table}[htb]
    \setlength{\tabcolsep}{3pt}
    \renewcommand{\arraystretch}{1.2}
    \begin{tabularx}{\linewidth}{llXX}
        Parameter &  &Device (iv) & Device (v) \\
        \hline\hline
        \multicolumn{4}{l}{\itshape Optical modes} \\ \hline
        $\Ddiff/2\pi$  & (\si{\hertz}) & $\phantom{-}1.13\times 10^{13}$ & $\phantom{-}1.97\times 10^{13}$ \\
        $\ga/2\pi$     & (\si{\hertz}) & $\phantom{-}1.00\times 10^{9}$  & $\phantom{-}6.00\times 10^{8}$  \\
        $\tka/2\pi$    & (\si{\hertz}) & $\phantom{-}1.00\times 10^{13}$ & $\phantom{-}1.00\times 10^{13}$ \\
        $\tkd/2\pi$    & (\si{\hertz}) & $\phantom{-}3.25\times 10^{9}$  & $\phantom{-}1.08\times 10^{9}$  \\
        $\lambda/2\pi$ & (\si{\hertz}) & $\phantom{-}4.09\times 10^{11}$   & $\phantom{-}4.09\times 10^{11}$   \\
        \hline
        \multicolumn{4}{l}{\itshape Mechanical mode} \\ \hline
        $\Om/2\pi$     & (\si{\hertz}) & $\phantom{-}1.3\times 10^6$     & $\phantom{-}1.3\times 10^6$     \\
        $\gam/2\pi$    & (\si{\hertz}) & $\phantom{-}9.3\times 10^{-3}$  & $\phantom{-}9.3\times 10^{-3}$  \\
        $Q_\m$         &               & $\phantom{-}1.4\times 10^8$     & $\phantom{-}1.4\times 10^8$     \\
        \hline
        \multicolumn{4}{l}{\itshape Single-photon optomechanical couplings} \\ \hline
        $\gwa/\Om$     &               & $\phantom{-}0.065$              & $\phantom{-}0.65$               \\
        $\gwd/\Om$     &               & $-0.14$                         & $-1.4$                          \\
        $\gka/\Om$     &               & $\phantom{-}0.060$              & $\phantom{-}0.60$               \\
        $\gkd/\Om$     &               & $\phantom{-}2.1\times 10^{-4}$  & $\phantom{-}4.2\times 10^{-4}$  \\
        \hline
        \multicolumn{4}{l}{\itshape Effective optical ``$-$'' mode } \\ \hline
        $\tkm/\Om$     &               & $\phantom{-}0.25$               & $\phantom{-}0.05$               \\
        $\gw[-]/\Om$   &               & $-0.14$                         & $-1.4$                          \\

        $\gw[-]/\tkm$  &               & $-0.56$                         & $-28$                           \\

        \hline\hline
    \end{tabularx}
    \caption{\label{tab:params2}
        Experimentally achievable devices in the effective resolved-sideband regime ($\tkm/\Om < 1$), thanks to the condition $\Ddiff = \kdiff \lambda/\sqrt{\tka\tkd}$, going toward both the strong ($\gw[-] > \tkm$) and ultrastrong ($\gw[-] > \Om$) coupling regimes. See Table~\ref{tab:params} for the description of the parameters. Like for Devices (ii) and (iii), the effective dissipative coupling $\gk[-]$ is negligible, $\gk[-] \ll \gw[-]$, see Sec.~\ref{sec:model:mech:micro}.
    }
\end{table}

Compared to the previous sections, where the device parameters were based on Table~\ref{tab:params}, we now chose more realistic optomechanical couplings for modes $a$ and $d$ for experimental microcavity implementations like in Ref.~\cite{Manjeshwar2023May} (Device (i)). Concretely, comparing to Ref.~\cite{Manjeshwar2023May} we only decrease the optomechanical couplings by a factor 10 for Device (iv) and keep the same values as in Device (i) for Device (v). Note that we further decreased $\gkd$ to reflect the smaller value of $\tkd$ in Devices (iv) and (v) compared to Device (i).
Despite the stronger optomechanical coupling, the modeling from Sec.~\ref{sec:model:optical} remains valid for Devices (iv) and (v),  in particular for the calculation of the effective optical modes from Sec.~\ref{sec:model:optical:langevin} since $\tka, \tkd \gg \gwa, \gwd, \gka, \gkd$. We can therefore make quantitative statements about the relevant effective coupling parameters, given in the last section of Table~\ref{tab:params2}.

Here, we find that Device (iv) is close to reaching both the effective strong coupling regime, $|\gw[-]| > \tkm$, and the effective ultrastrong coupling regime, $|\gw[-]| > \Omega$. Indeed, we see in the last rows of Table~\ref{tab:params2} that $|\gw[-]| =0.56 \tkm$ and $|\gw[-]| =0.14 \Omega$, compared to the devices in Table~\ref{tab:params}, where these parameters typically differ by several orders of magnitude. The situation is even more favorable for Device (v) reaching both the effective strong coupling regime, $|\gw[-]| = 28 \tkm$, and the effective ultrastrong coupling regime, $|\gw[-]| =1.4 \Omega$. The required parameters of the mirror mode, $\omd$ and $\kd$, can be realized by engineering the photonic crystal pattern and the right-hand-side cavity loss rate, $\ga$, is achievable with distributed Bragg reflectors. Our work thus predicts that reaching simultaneously the resolved-sideband, strong and ultrastrong coupling regimes is within reach of optical microcavities with a frequency-dependent mirror such as a photonic crystal, paving the way for the experimental exploration of the nonlinear regime of cavity quantum optomechanics.

\begin{figure}[t]
    \includegraphics[width=\linewidth]{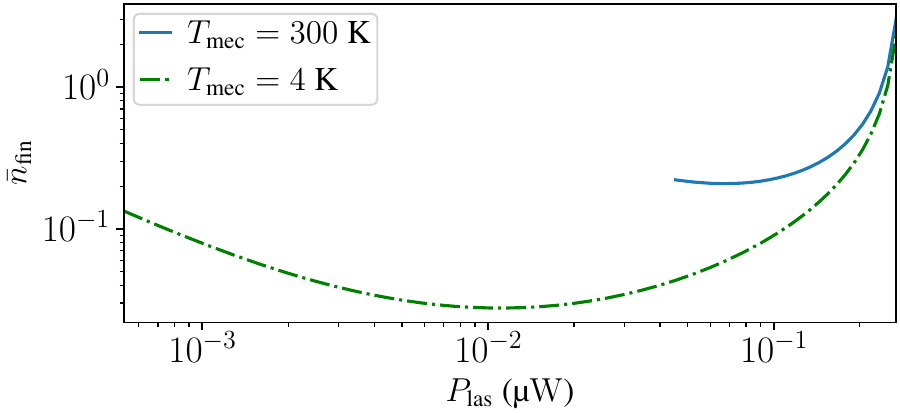}
    \caption{\label{fig:cooling_device_iv}
        Steady-state phonon number in the mechanical fluctuations (Eq.~\eqref{Neff}) as function of the laser power at $\tDm = \Om$ for Device (iv). The curves have been plotted on the laser power range where the device's dynamics can be linearized (Sec.~\ref{sec:linearized om}).}
\end{figure}

In general, the theoretical approach to analyze the \emph{dynamics} presented here, namely the linearization from Sec.~\ref{sec:linearized om}, does not apply for this strong-coupling regime.
However for strong enough laser powers, the dynamics of Device (iv) can be linearized since $\abs{\bc}^2 \gg \mean{\dc^\dagger\dc^{}}$, see also Appendix~\ref{app:validity linearization}. Hence, for those laser powers, we are able to show results for the phonon number that can be reached via optomechanical cooling, using the theoretical framework given in Sec.~\ref{sec:model}. The result is shown in Figure~\ref{fig:cooling_device_iv}, displaying the steady-state phonon number in the mechanical fluctuations  at $\tDm = \Om$, evidencing that ground-state cooling is possible with this device too (See Appendix~\ref{app:GS cooling} for the energy equipartition). The linearization is valid for a smaller range of laser powers at room temperature (solid blue line) than at cryogenic temperature (dashed-dotted green line) since the thermal fluctuations of the mechanics, which give rise to optical fluctuations due to the optomechanical couplings, increase with temperature. An analysis of the full nonlinear dynamics of Devices (iv) and (v) is postponed to future work.

\section{Conclusion}\label{sec:conclusion}

Our work highlights that optomechanical systems with a Fano mirror can reach an \emph{effective} resolved-sideband regime. Thanks to the strong interaction between the two optical modes, i.e., the Fano and the cavity modes, a significant reduction of the effective optical linewidth is achievable by matching the resonance frequencies and loss rates of these two modes.
Therefore sideband-based ground-state cooling of the mechanics is expected to become possible even for a microcavity with high optical losses $\tka \gg \Om$ and at room temperature.

We have established a complex but versatile theoretical model, taking into account both dispersive and dissipative optomechanical couplings for a cavity with one frequency-dependent mirror. The full analytical description makes it possible to investigate different effects separately, namely couplings not relevant to a chosen specific experimental setup can easily be set to zero. Furthermore, the cavity output spectrum is not trivially related to the noise spectrum of the mechanical resonator and our work gives the required expressions to make this inference.
We therefore expect this model to be a useful tool for all similar or analogous setups, where coupled optical modes interact with mechanical modes. This can be applied to various systems ranging from optomechanical cavities studied here to magnon modes coupled to mechanical deformation~\cite{Zhang2016Mar,Potts2021Sep}, molecular optomechanics~\cite{Zou2023Aug}, or cavity-enhanced chemistry etc., see e.g., Refs.~\cite{Li2022Apr, Simpkins2023Apra}.

Finally, we have demonstrated that this complex optomechanical system can be mapped to a canonical optomechanical setting with ideal coupling and loss parameters. Importantly, this shows that the advantage of strong single-photon coupling of microcavities and the designed reduced loss rates of effective optical modes due to a Fano mirror are a promising route for implementing both strong coupling and ultrastrong coupling in optomechanics, paving the way for nonlinear quantum optomechanics.

\acknowledgments

We thank Lei Du for helpful comments on the manuscript. We acknowledge financial support by the Knut och Alice Wallenberg stiftelse through project grant no. 2022.0090 (J.S., W.W.), the Wallenberg Center for Quantum Technology (A.C.) as well as through individual Wallenberg Academy fellowship grants (J.S., W.W.). Also funding from the Swedish Vetenskapsr\r{a}det via project numbers 2018-05061 (J.M., J.S.), 2019-04946 and 2019-00390 (W.W.), and the QuantERA project (C’MON-QSENS!) are gratefully acknowledged.

\appendix

\section{Dynamics}\label{app:dynamics}

In this appendix section, we provide derivations and technical background for the (linearized) Langevin equations and the resulting Lyapunov equations, which are used throughout the main paper to describe the dynamics of the optomechanical system.

\subsection{Derivation of the Langevin equations}\label{app:Langevin}

From the microscopic model described in Sec.~\ref{sec:model:mech:micro}, we derive the dynamics of the system
in the Heisenberg picture,
\begin{widetext}
\begin{subequations}\label{Heisenberg eqs}
    \begin{align}
        \dot{\hat{b}}_{\omega,\L}=\,& -i\omega \bwL - \sqrt{\frac{\ka}{\pi}}\left(1 + \frac{\gka}{\sqrt{2}\ka} \hat{q}\right)\hat{a}\label{db1w}- \sqrt{\frac{\kd}{\pi}}\left(1 + \frac{\gkd}{\sqrt{2}\kd} \hat{q}\right)\hat{d},\\
        \dot{\hat{b}}_{\omega,\R}=\,& -i\omega \bwR - \sqrt{\frac{\ga}{\pi}}\hat{a},\label{db2w}\\
        \dot{\hat{a}}=\,& -i\oma \hat{a} -i\lambda \hat{d} + i \gwa\sqrt{2}\hat{q}\hat{a}   +  \sqrt{\frac{\ka}{\pi}}\left(1 + \frac{\gka}{\sqrt{2}\ka} \hat{q}\right)\int \dl \omega\bwL+  \sqrt{\frac{\ga}{\pi}}\int \dl \omega\bwR ,\\
        \dot{\hat{d}}=\,& -i\omd \hat{d}-i\lambda \hat{a} + i \gwd\sqrt{2}\hat{q}\hat{d}  +  \sqrt{\frac{\kd}{\pi}}\left(1 + \frac{\gkd}{\sqrt{2}\kd} \hat{q}\right)\int \dl \omega\bwL,\\
        \dot{\hat{q}}=\,& \Om \hat{p},\\
        \dot{\hat{p}}=\,& -\Om \hat{q} + \sum_c \left(\gw\sqrt{2}\hat{c}^\dagger \hat{c} - i \frac{\gk}{\sqrt{2\pi\kc}}\int \dl \omega (\hat{c}^\dagger \bwL^{} - \bwL^\dagger \hat{c}^{})\right).
    \end{align}
\end{subequations}
\end{widetext}
We formally integrate Eqs.~\eqref{db1w} and \eqref{db2w} with respect to an initial time
$t_0 < t$, and insert the obtained expression in the other equations of the system \eqref{Heisenberg eqs}. Defining the input fields
\begin{equation}
    \bin[,\mu](t) = \frac{1}{\sqrt{2\pi}}\int\dl\omega \bw(t_0)\e^{-i\omega(t - t_0) },
\end{equation}
with $\mu = \L,\R$, and following the same procedure as in Ref.~\cite{Gardiner1985Jun}, we obtain the Langevin equations Eqs.~\eqref{Langevin eqs}.
We have added the terms $-\gam \hat{p}$ and $\sqrt{\gam}\hat{\xi}$ in the evolution of $\hat{p}$ to take into account the thermal mechanical noise, assuming that $\gam \ll \Om, \gwa, \gka, \ga, \ka, \kd$.

The output fields are defined by taking a reference time in the future $t_1 > t$
\begin{equation}
    \bout[,\mu](t) = \frac{1}{\sqrt{2\pi}}\int\dl\omega \bw(t_1)\e^{-i\omega(t - t_1) }.
\end{equation}
Comparing the formal integration of Eqs.~\eqref{db1w} with respect to $t_0$ and $t_1$ gives the input-output relations Eqs.~\eqref{input-output}.

\subsection{Validity of the linearization}\label{app:validity linearization}

For the linearized dynamics described in Sec.~\ref{sec:linearized om} to be valid, two key conditions need to be fulfilled.
First, we have assumed the existence of a well-defined stable classical steady-state solution of the non-linear system of equations \eqref{ss}, which can be verified by applying the Routh-Hurwitz criterion, see below. Second, we have neglected terms at the second order in the fluctuations in the Langevin equations \eqref{Langevin lin eqs} and it needs to be guaranteed that the neglected terms are indeed sufficiently small. In the following, we do not discuss Device (i) since the linearization for this device is valid for all the considered detunings due to the large effective optical losses, $\tkm \gg \gw, \gk, \Om$, see Table~\ref{tab:params}.

\subsubsection{Stability of the system}\label{app:stability}

We apply the Routh-Hurwitz criterion to determine whether the system, described by the  Langevin equations \eqref{Langevin eqs}, is stable for the studied parameters.
Focusing on the equation for the quadratures and following the procedure described in Ref.~\cite{DeJesus1987Jun}, we construct the sequence $(r_k)_{0\le k \le 6}$ based on the characteristic polynomial of $A$ (Eq.~\eqref{A}), $\det(A - X I) = \sum_{k=0}^6 a_{6 - k} X^k$.
If all the elements of this sequence have the same sign then the system is stable.
The elements are defined in the following way, $r_0 = T_0$, $r_1 = T_1$ and $r_k = T_k/T_{k-1}$ for $k = 2, ..., 6$ with $T_0 = a_0$, $T_1 = a_1$ and $T_{k>1} = \det{M_k}$.
Here, $M_k$ is the $k\times k$ matrix with coefficients $(M_k)_{ij} = a_{2i - j}$ and we take $a_{2i - j} = 0$ if $2i - j$ is not between 0 and 6. In particular, $r_0 = 1$, $r_1 = 2(\ga + \ka + \kd) + \gam$ and $r_6 = a_6 = \det(A)$ but the full analytical expressions of the other coefficients is rather complex and not very informative.
However, by evaluating numerically $r_{k>1}$ we can find that Devices (ii) and (iii) have instabilities (see hatched regions in relevant plots of the main paper) which coincide very well with the places at which $\Neff$ becomes negative.

\begin{figure}[htb]
    \includegraphics[width=\linewidth]{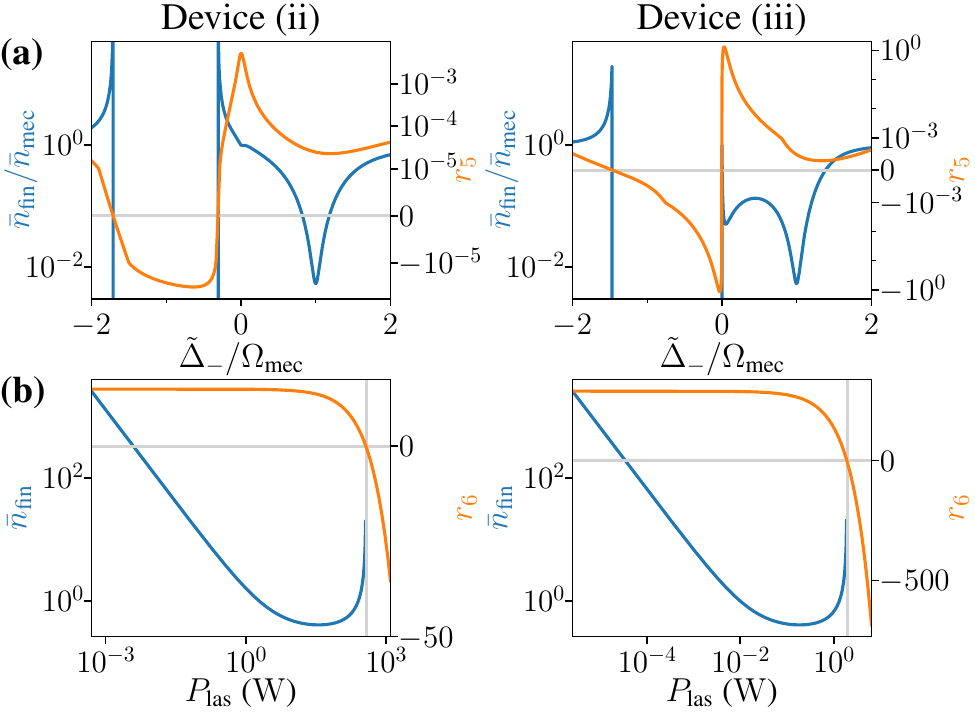}
    \caption{\label{fig:stability}
        Steady-state phonon number $\Neff$ (in blue, left axis) and relevant stability condition (in orange, right axis) as a function of \textbf{(a)} the detuning $\tDm$ and \textbf{(b)}  the laser power $\Plas$ for Devices (ii) and (iii). See Table \ref{tab:params} for the other parameters.
    }
\end{figure}

In particular, the instability regions in Figs.~\ref{fig:dTs}, \subfigref{fig:dOm and Gopt}{a} and \ref{fig:weak coupling} as a function of the detuning $\tDm$ for a fixed laser power coincide with negativities of $r_5$, see Fig.~\subfigref{fig:stability}{a}.
However, when the system is driven on the red sideband, at $\tDm = \Om$, the relevant criterion, namely the first one to become negative, for the stability of the system as a function of the laser power is $r_6 = \det(A) > 0$, as illustrated in Fig.~\subfigref{fig:stability}{b}. Similar results are obtained for the other devices.

\subsubsection{Negligibility of the second order in the fluctuations}\label{app:linearization}

\begin{figure}[htb]
    \includegraphics[width=\linewidth]{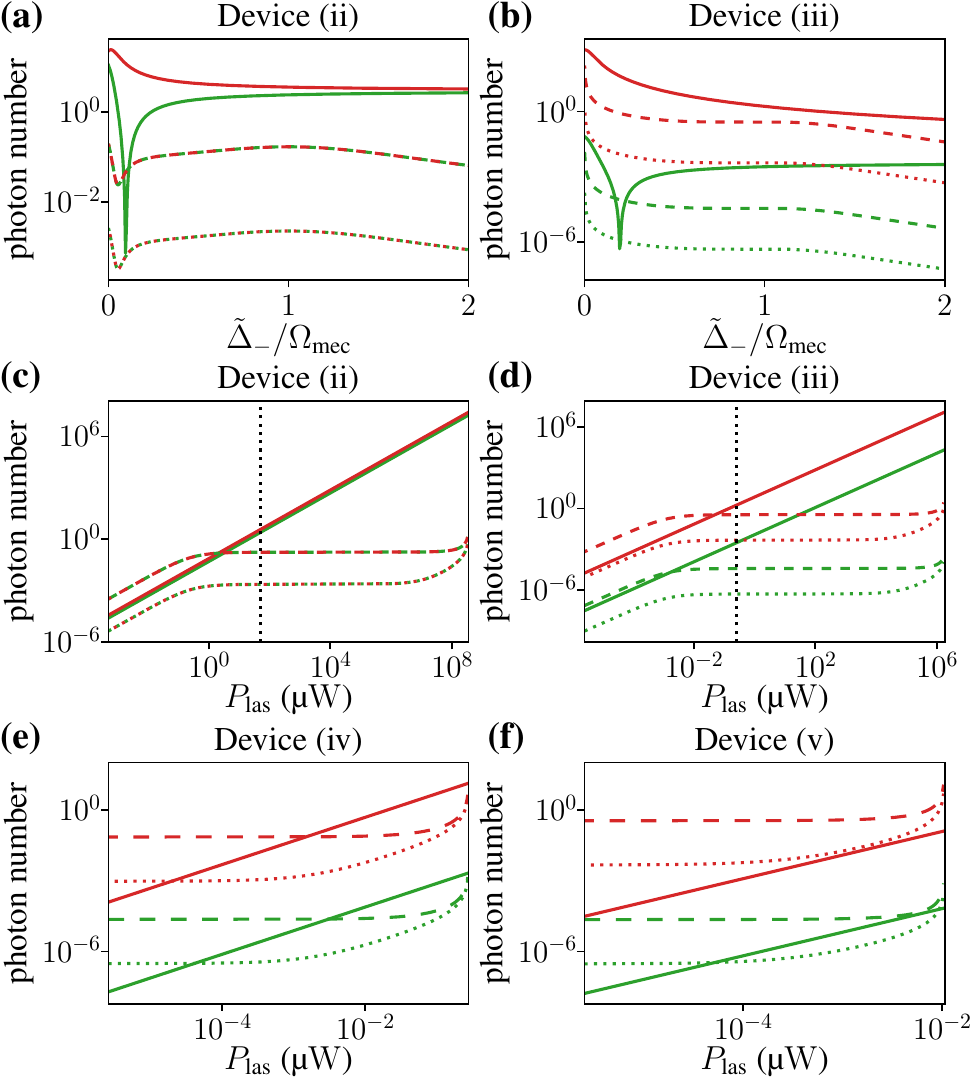}
    \caption{\label{fig:lin validity}
        Steady-state photon numbers, for the cavity mode in green and the mirror mode in red, in the classical steady-state $\abs{\bc}^2$ (solid lines) and in the fluctuations $\mean{\dc^\dagger\dc^{}}$ (dashed lines at $T_\m = 300$ \si{\kelvin} and dotted lines at $T_\m = 4$ \si{\kelvin}) as a function of \textbf{(a)},\textbf{(b)} the detuning $\tDm$ and \textbf{(c-f)} the laser power $\Plas$ (at $\tDm = \Om$) for the device indicated on top of each panel. The dotted vertical black lines in panels (c) and (d) indicate the laser powers used in all the figures where $\Plas$ is not on the x-axis. See Table~\ref{tab:params} for the other parameters of Devices (ii) and (iii) and Table~\ref{tab:params2} for the parameters of Devices (iv) and (v).
    }
\end{figure}

We compare the fluctuations in the photon numbers, $\mean{\dc^\dagger\dc^{}}$, with the average photon numbers in the classical steady state, $\abs{\bc}^2$ from Eq.~\eqref{ss}, for both the cavity mode ($c=a$ in green) and mirror mode ($c=d$ in red) in Fig.~\ref{fig:lin validity}.
The photon number fluctuations $\mean{\dc^\dagger\dc^{}}$ are obtained by numerically solving the steady-state Lyapunov equations for the second-order moments, see Appendix \ref{app:Lyapunov}.
While $\abs{\bc}^2$ is independent of the temperature $T_\m$, since thermal fluctuations average to zero, $\mean{\dc^\dagger\dc^{}}$ decreases by several orders of magnitude when going from room temperature (dashed lines) to cryogenic temperature (dotted lines).
Indeed, the thermal fluctuations in the mechanical resonator create optical fluctuations due to the optomechanical coupling; therefore, having a colder mechanical environment reduces these thermal fluctuations.
Figs.~\subfigref{fig:lin validity}{a,b} correspond to the positive detuning part of Fig.~\ref{fig:weak coupling} and it shows that apart from around $\ba\approx 0$, which happens at $\omL = \omd - \sqrt{\tkd/\tka}\lambda$, see Eq.~\eqref{ss}, the linearization is valid for Devices (ii) and (iii).

Figs.~\subfigref{fig:lin validity}{c-e} show that, at $\tDm = \Om$, the linearization is valid if the laser power is sufficiently large for Devices (ii)-(iv), which is the case in all figures presented in this paper. The laser powers used in most figures, see Table \ref{tab:params}, for Devices (ii) and (iii) are indicated by the vertical dotted black line in Fig.~\subfigref{fig:lin validity}{b}. However, the linearization for Device (v), in Fig.~\subfigref{fig:lin validity}{f}, is never valid, which is expected since this device is in the strong coupling regime, $\gw[-] > \tkm$.

\subsection{Solution of the linearized Langevin equations}\label{app:solution Langevin}

In the frequency domain, using the convention  $\hat{f}[\omega] = \int_{-\infty}^{+\infty} \dl t \e^{i\omega t}\hat{f}(t)$ for the Fourier transform, the linearized Langevin equations \eqref{Langevin lin eqs} become

    \begin{align}
        \chi_a^{-1}[\omega] \da =\,&i\sqrt{2}\tga\dq -i\tG\dd + \sqrt{2\tka}\ainL + \sqrt{2\ga}\ainR,\nonumber\\\nonumber
        \chi_d^{-1}[\omega]\dd =\,&i\sqrt{2}\tgd\dq -i\tG\da  + \sqrt{2\tkd}\ainL ,\\\nonumber
        \deltp =\,& -i\frac{\omega}{\Om}\dq ,\\\nonumber
        \chi_{\m,0}^{-1}[\omega]\dq =\,& \sqrt{2} \sum_{c}(\tgm \dc^\dagger +\tgm^*\dc^{})+ \sqrt{\gam}\hat{\xi} \\
        &+ c_X\sqrt{2}\XinL + c_P\sqrt{2}\PinL.\label{TF Langevin lin eqs}
    \end{align}
The susceptibilities $\chi_a$, $\chi_d$, and $\chi_{\m,0}$ are given in Sec.~\ref{sub:frequency_solution}. From these equations, we obtain
\begin{equation}
    \dc = \sqrt{2}C^c_q[\omega]\dq + C^c_\L[\omega]\ainL + C^c_\R[\omega]\ainR, \label{dc}
\end{equation}
with the coefficients $C^c$ defined in Eqs.~\eqref{C_mu} and \eqref{Cq_c}.
Then we get Eqs.~\eqref{solution Langevin:q} and \eqref{Xopt} by putting Eq.~\eqref{dc} in the last equation of \eqref{TF Langevin lin eqs}.

Furthermore, we can write each quadrature $\hat{Q}$, with $Q = X_a, P_a, X_d, P_d, q, p$, as
\begin{equation}\label{dQ}
    \dQ[\omega] = \sum_{\eta} c^Q_\eta[\omega] \hat{\eta}[\omega],
\end{equation}
where we are summing over the input noises, $\hat{\eta} = \XinL, \PinL, \XinR, \PinR, \hat{\xi}$, and we have defined the coefficients

{\allowdisplaybreaks
\begin{align}
    c^q_{X_{\text{in,}\mu}}[\omega] =\,& \chieff[\omega]\sum_{c}\left(\tgm^* C^c_\mu[\omega]+ \tgm C^c_\mu[-\omega]^*\right) \nonumber\\\nonumber
    & + \delta_{\mu,\L} \chieff[\omega] c_X \sqrt{2} ,
    \\\nonumber
    c^q_{P_{\text{in,}\mu}}[\omega] =\,&\chieff[\omega]\sum_{c}i\left(\tgm^* C^c_\mu[\omega]-\tgm C^c_\mu[-\omega]^* \right) \nonumber\\\nonumber
    &+ \delta_{\mu,\L}\chieff[\omega]c_P \sqrt{2},
    \\\nonumber
    c^q_\xi[\omega] =\,& \chieff[\omega]\sqrt{\gam},
    \\
    c^p_\eta[\omega] =\,& -i\frac{\omega}{\Om}c^q_\eta[\omega],\nonumber\\
    c^{X_c}_{X_{\text{in,}\mu}}[\omega]  =\,& (C^c_q[\omega] + C^c_q[-\omega]^*)c^q_{X_{\text{in,}\mu}}[\omega] + \frac{C^c_\mu[\omega] + C^c_\mu[-\omega]^*}{2},
    \nonumber\\\nonumber
    c^{X_c}_{P_{\text{in,}\mu}}[\omega]  =\,& (C^c_q[\omega] + C^c_q[-\omega]^*)c^q_{P_{\text{in,}\mu}}[\omega] + i\frac{C^c_\mu[\omega] - C^c_\mu[-\omega]^*}{2},
    \\\nonumber
    c^{X_c}_{\xi}[\omega]  =\,& (C^c_q[\omega] + C^c_q[-\omega]^*)c^q_{\xi}[\omega],
    \\\nonumber
    c^{P_c}_{X_{\text{in,}\mu}}[\omega]  =\,& i(C^c_q[-\omega]^*\!-C^c_q[\omega])c^q_{X_{\text{in,}\mu}}[\omega] +i \frac{C^c_\mu[-\omega]^*\! - C^c_\mu[\omega]}{2},
    \\\nonumber
    c^{P_c}_{P_{\text{in,}\mu}}[\omega]  =\,& i(C^c_q[-\omega]^*\!-C^c_q[\omega])c^q_{P_{\text{in,}\mu}}[\omega] + \frac{C^c_\mu[\omega] + C^c_\mu[-\omega]^*}{2},
    \\
    c^{P_c}_{\xi}[\omega]  =\,& i(C^c_q[-\omega]^* - C^c_q[\omega])c^q_{\xi}[\omega].\label{coefs}
\end{align}}

\subsection{Evolution of the second-order moments}\label{app:Lyapunov}

Due to the linearization in Sec.~\ref{sec:linearized om}, the system is Gaussian Therefore, the evolution of the second-order moments of the quadratures can be put in the form of a Lyapunov equation
\begin{equation}\label{Lyapunov equation}
    \diff{V}{t} = A V+V A^T+ B,
\end{equation}
for the covariance matrix $V$ of the quadratures.
The elements of the covariance matrix $V$ are defined as
\begin{equation}\label{cov matrix elements}
    V_{ij}= \dfrac{1}{2} \mean{\{ Y_i, Y_c\}}  - \mean{Y_i} \mean{Y_c},
\end{equation}
with $\vec{Y} = \left( \dXa, \dPa, \dXd, \dPd, \dq, \deltp  \right)$. We have defined the optical quadratures $\dXa = (\da + \da^\dagger)/\sqrt{2}$, $\dPa = (\da - \da^\dagger)/i\sqrt{2}$, $\dXd = (\dd + \dd^\dagger)/\sqrt{2}$ and $\dPd = (\dd - \dd^\dagger)/i\sqrt{2}$. The expressions of the matrices $A$ and $B$,
\begin{widetext}
    \begin{align}\label{A}
        A &= \begin{bmatrix}
            -\tka - \ga & \tDa  &-\sqrt{\tka\tkd} & \lambda &-2\Im(\tga) & 0\\
            -{\tDa} & -\tka - \ga  & -\lambda & -\sqrt{\tka\tkd} & 2  \Re(\tga) & 0\\
            -\sqrt{\tka\tkd} & \lambda &-{\tkd} & {\tDd}& - 2\Im(\tgd)  & 0  \\
            -\lambda & -\sqrt{\tka\tkd} &  -{\tDd} & -{\tkd} &2\Re(\tgd)  & 0\\
            0 & 0  & 0 & 0& 0 & \Om \\
            2 \Re(\tgma) &2 \Im(\tgma)  & 2 \Re(\tgmd) & 2 \Im(\tgmd) & -\Om & -{\gam}\\
        \end{bmatrix}, \\ \nonumber
        B &=\begin{bmatrix}
            \tka + \ga & 0 & \sqrt{\tka\tkd} & 0 &0 & c_X\sqrt{\tka} \\
            0 & \tka + \ga & 0 & \sqrt{\tka\tkd} & 0 &  c_P\sqrt{\tka} \\
            \sqrt{\tka\tkd} & 0 & \tkd & 0 & 0 &c_X\sqrt{\tkd} \\
            0 & \sqrt{\tka\tkd} & 0 & \tkd & 0 & c_P\sqrt{\tkd}\\
            0 & 0 & 0 & 0 & 0 & 0 \\
            c_X\sqrt{\tka} & c_P\sqrt{\tka} &  c_X\sqrt{\tkd} &  c_P\sqrt{\tkd}& 0 &c_P^2 + c_X^2 +
            \gam\left(2 {\Nm} + 1\right)
        \end{bmatrix}.
    \end{align}
\end{widetext}
are obtained from the Langevin equations~\eqref{Langevin lin eqs} and the correlation functions of the noise, Eqs.~\eqref{correlations_xi} and \eqref{correlations_a_in}.
Solving numerically the Lyapunov equation \eqref{Lyapunov equation} for the steady-state, $A \bar{V}+\bar{V} A^T+ B = 0$,  gives access to, among other quantities, the phonon number in the mechanical fluctuations, Eq.~\eqref{Neff},
\begin{equation}
    \Neff = \frac{1}{2}(\bar{V}_{55} + \bar{V}_{66} - 1).
\end{equation}

\section{Detailed analytical expressions for the mean optical response}\label{app:mean opt}

The optical response of the optomechanical system is discussed in Sec.~\ref{sec:OM:optical_response}. Here, we provide explicit analytical expressions, which are not given in Sec.~\ref{sec:OM:optical_response}.

{\allowdisplaybreaks
The derivative of the intensity transmission $T$ in Eq.~\eqref{dTdq} can be explicitly expressed as
\begin{align}
    \diffp{T}{\tDa} &= -\diffp{D}{\tDa}\frac{T}{D},\\\nonumber
    \diffp{T}{\tDd} &= -\diffp{D}{\tDd}\frac{T}{D} +\frac{8\ga}{D}\left(\tka\tDd - \sqrt{\tka\tkd}\lambda\right),\\\nonumber
    \diffp{T}{\tka} &= -\diffp{D}{\tka}\frac{T}{D} +\frac{4\ga\tDd}{D}\left(\tDd - \sqrt{\frac{\tkd}{\tka}}\lambda\right),\\
    \diffp{T}{\tkd} &= -\diffp{D}{\tkd}\frac{T}{D} -\frac{4\ga\lambda}{D}\left(\sqrt{\frac{\tka}{\tkd}}\tDd - \lambda\right),\nonumber
\end{align}
with
\begin{align}
    \diffp{D}{\tDa} =\,& 2\tDd(\tka\tkd - \lambda^2) - 4\lambda\tkd\sqrt{\tka\tkd} + 2(\tDd^2 + \tkd^2)\tDa,\nonumber\\ \nonumber
    \diffp{D}{\tDd} =\,& 2\tDa(\tka\tkd - \lambda^2) - 4\lambda(\tka + \ga)\sqrt{\tka\tkd} \\\nonumber
    &+ 2(\tDa^2 + (\tka + \ga)^2)\tDd,\\\nonumber
    \diffp{D}{\tka} =\,& 2\!\left[\tDd - \sqrt{\frac{\tkd}{\tka}}\lambda\right]\!\!\left[\tDa\tkd + \tDd(\tka + \ga) - 2\lambda\sqrt{\tka\tkd}\right],\\ \nonumber
    \diffp{D}{\tkd} =\,& 2\!\left[\tDa - \sqrt{\frac{\tka}{\tkd}}\lambda\right]\!\!\left[\tDa\tkd + \tDd(\tka + \ga) - 2\lambda\sqrt{\tka\tkd}\right]\\
    &+2\ga\left(\lambda^2 + \ga\tkd - \tDa\tDd\right).
\end{align}
}

\section{Power spectra}\label{app:power spectra}
In this appendix section, we compute the power spectra of the mechanical position fluctuations and some of the optical quadratures. In the following, the power spectrum of an operator $\dQ$ is denoted $S_Q[\omega]$, with
\begin{align}
    S_Q[\omega] &\equiv \int_{-\infty}^{+\infty} dt \e^{i\omega t}\mean{\dQ(t)\dQ(0)}\nonumber \\
    &= \int_{-\infty}^{+\infty} \frac{d\omega'}{2\pi}\mean{\dQ[\omega]\dQ[\omega']}.\label{power spectrum}
\end{align}
and we will also use the notation
\begin{equation}
    S_{Q_1 Q_2}[\omega] = \frac{1}{2}\int_{-\infty}^{+\infty} \frac{\dl \omega'}{2\pi} \mean{\dQ_1[\omega]\dQ_2[\omega'] +\dQ_2[\omega]\dQ_1[\omega']},
\end{equation}
for correlators between two operators $\dQ_1$ and $\dQ_2$.\\

For measurements such as homodyne detection, the relevant spectrum is the symmetrized spectrum $\frac{1}{2}(S_Q[\omega] + S_Q[-\omega])$. Using Eq.~\eqref{dQ}, we get
\begin{equation}
        S_Q[-\omega] =\sum_{\eta,\eta'} \int_{-\infty}^{+\infty} \frac{d\omega'}{2\pi} c^Q_\eta[-\omega] c^Q_{\eta'}[\omega']  \mean{ \hat{\eta}[-\omega]\hat{\eta'}[\omega']},
\end{equation}
and, given the noise correlations functions Eqs.~\eqref{correlations_a_in} and \eqref{correlations_xi}, we have $\mean{ \hat{\eta}[-\omega]\hat{\eta'}[\omega']} \propto \delta(\omega' - \omega)$. Therefore, $S_Q[-\omega] = S_Q[\omega]$ and the symmetrized spectrum is also given by Eq.~\eqref{power spectrum}.

\subsection{Mechanical position spectrum}\label{app:position spectrum}

We first compute the mechanical position power spectrum, $S_q[\omega]$.
Using Eq.~\eqref{solution Langevin:q} and the noise correlation functions, Eqs.~\eqref{correlations_xi} and \eqref{correlations_a_in}, in the frequency domain,
we get Eq. \eqref{Sq}.
In the limit of a purely dispersive optomechanical coupling, $\gka\to 0$, we recover Eq.~(D10) from \cite{Monsel2021Jun}.
The effective susceptibility, which appears in Eq. \eqref{Sq}, is defined by Eq.~\eqref{susceptibility} and the related coefficients $C_\mu^{a/d}$ by Eqs.~\eqref{C_mu}.
The phonon number in the mechanical fluctuations, Eq.~\eqref{Neff}, can also be computed from the position spectrum \cite{Genes2008Mar} since
\begin{align}\label{dq2 dp2}
    \mean{\dq^2} &= \int_{-\infty}^{+\infty} \frac{\dl \omega}{2\pi} S_q[\omega],\\\nonumber
    \mean{\deltp^2} &= \int_{-\infty}^{+\infty} \frac{\dl \omega}{2\pi} \left(\frac{\omega}{\Om}\right)^2S_q[\omega].
\end{align}
Note that in practice, these integrals only need to be evaluated on a frequency range $[0, \Omega_\text{max}]$, with the cutoff frequency $\Omega_\text{max}$ one or two orders of magnitude larger than $\Om$, since the integrands are even functions of $\omega$ and $S_q[\omega]$ is sharply peaked at the effective mechanical frequency.

\subsection{Output light spectrum}\label{app:output spectrum}

As a next step, we compute the power spectra, as defined in Eq.~\eqref{power spectrum}, of the quadratures of the light leaking out of the cavity, i.e., $S_{X_{\Out,\mu}}$ and $S_{P_{\Out,\mu}}$, where $\Xout = (\aout^{} + \aout^\dagger)/\sqrt{2}$ and $\Pout = (\aout^{} - \aout^\dagger)/i\sqrt{2}$ are the position and momentum quadratures of the output light. The input-output relation for these quadratures is
\begin{align}\nonumber
    \!\!\XoutL &=  \XinL - \!\sqrt{2\tka}\dXa- \!\sqrt{2\tkd}\dXd  +\sqrt{2}c_P \dq,\!\\\nonumber
    \PoutL &=  \PinL  - \sqrt{2\tka}\dPa - \sqrt{2\tkd}\dPd+\sqrt{2}c_X \dq,\\\nonumber
    \!\!\XoutR &=  \XinR - \!\sqrt{2\ga}\dXa,\!\\
    \PoutR &=  \PinR  - \sqrt{2\ga}\dPa,
\end{align}
see Eq.~\eqref{input-output lin}. Therefore,
we get the output spectra given in Eq.~\eqref{output spectra}.
Using the solutions of the Langevin equations, Eq.~\eqref{dQ} and the noise correlation functions, we compute the fluctuation spectra
\begin{widetext}
    \begin{align}
    S_{Q}[\omega]
     =\,& \frac{1}{2} \sum_{\mu}\abs*{c_{{X}_{\text{in,}\mu}}^{Q}[\omega] - ic_{{P}_{\text{in,}\mu}}^{Q}[\omega]}^2 + (2\Nm + 1)\abs*{c_{\hat{\xi}}^{Q}[\omega]}^2\\
    S_{Q_1 Q_2}[\omega] =\,
    & \frac{1}{2}\sum_{\mu}\left[ \Re \left\{\left(c_{{X}_{\text{in,}\mu}}^{Q_1}[\omega] - ic_{{P}_{\text{in,}\mu}}^{Q_1}[\omega]\right)\left(c_{{X}_{\text{in,}\mu}}^{Q_2}[\omega] - ic_{{P}_{\text{in,}\mu}}^{Q_2}[\omega]\right)^*   \right\}\right]\nonumber
    + \left(2\Nm + 1\right)\Re\left\{c_{\hat{\xi}}^{Q_1}[\omega]c_{\hat{\xi}}^{Q_2}[-\omega]\right\}  ,
\end{align}
\end{widetext}
with $Q, Q_1, Q_2 \in\{ X_a, P_a, X_d, P_d, q\}$. The coefficients $c^{P_c}_\eta$ are defined in Eqs.~\eqref{coefs} and note that $c_{\eta}^{Q}[-\omega] = c_{\eta}^{Q}[\omega]^*$.

In many cases, including all devices considered here, the part of the noise power spectrum coming from the optical environments (vacuum noise) is negligible compared to the thermal noise from the mechanical environment. As a consequence, we have
\begin{align}
    S_Q[\omega] &\simeq  2 \Nm\abs*{c_{\hat{\xi}}^{Q}[\omega]}^2, \\\nonumber
    S_{Q_1Q_2}[\omega] &\simeq  2 \Nm\Re\left\{c_{\hat{\xi}}^{Q_1}[\omega]c_{\hat{\xi}}^{Q_2}[-\omega]\right\},
\end{align}
and, in particular, $S_q[\omega] \simeq 2\gam\Nm \abs{\chi_\m^\eff[\omega]}^2$.
The expression for $c_{\hat{\xi}}^{Q}[\omega]$ always contains the factor $\sqrt{\gam}  \chi_\m^\eff[\omega]$, see Eqs.~\eqref{coefs}.
Therefore, $S_{X_{\text{out},\L}}[\omega]$ and $S_{P_{\text{out},\L}}[\omega]$ can be approximated by $\gam  \abs{\chi_\m^\eff[\omega]}^2$ times a frequency-dependent factor and they hence exhibit a peak at the effective mechanical frequency.

\begin{figure}[b]
    \includegraphics[width=\linewidth]{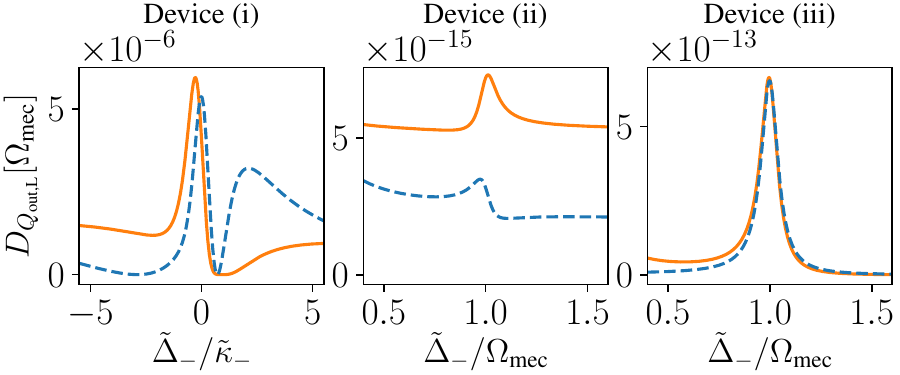}
    \caption{\label{fig:Dfactor}
        Factor $D_{Q_{\Out,\L}}[\Om]$ for $Q=X$ (solid orange) and $Q=P$ (dashed blue) for Devices (i-iii) as a function of the detuning $\tDm$. See Table~\ref{tab:params} for the parameters.
    }
\end{figure}
However, in all generality, they do not give direct access to $\mean{\dq^2}$ by simple integration over a range of values of $\omega$, $[0, \Omega_\text{max}]$, relevant for mechanical features, like in Eq.~\eqref{dq2 dp2}, because of the frequency-dependent prefactors. We can nonetheless make $S_q[\omega]$ appear in the expression of all the $S_Q[\omega]$ and $S_{Q_1Q_2}[\omega]$, such that
\begin{equation}
     S_{Q_{\Out,\mu}}[\omega] = S_{Q_{\In,\mu}}[\omega] + D_{Q_{\Out,\mu}}[\omega]S_q[\omega],
\end{equation}
with $Q = X, P$, $\mu = \L, \R$ and where we have defined
\begin{align}
D_{X_{\Out,\L}}[\omega] =\, &\abs*{\sqrt{2}c_P + \sum_c \sqrt{2\tk} (C^c_q[\omega] + C^c_q[-\omega]^*)}^2,\nonumber\\
D_{P_{\Out,\L}}[\omega] =\, &\abs*{\sqrt{2}c_X + \sum_c \sqrt{2\tk} i(C^c_q[-\omega]^* - C^c_q[\omega])}^2,\nonumber\\
D_{X_{\Out,\R}}[\omega] =\,
& 2\ga\abs*{C^a_q[\omega] + C^a_q[-\omega]^*}^2, \nonumber\\
D_{P_{\Out,\R}}[\omega] =\,
& 2\ga\abs*{C^a_q[\omega] - C^a_q[-\omega]^*}^2.
\end{align}
We can then compute $\mean{\dq^2}$ as
\begin{align}
 \mean{\dq^2}
  &= 2\int_{0}^{\Omega_\text{max}} \frac{\dl \omega}{2\pi}\frac{S_{Q_{\Out,\mu}}[\omega] - S_{Q_{\In,\mu}}[\omega]}{D_{Q_{\Out,\mu}}[\omega]}.
\end{align}
Note that for the devices we consider, the frequency dependence of $D_{Q_{\Out,\mu}}[\omega]$ for $\omega \in [0, \Omega_\text{max}]$ is weak, especially for Device (i), such that $D_{Q_{\Out,\mu}}[\omega] \simeq D_{Q_{\Out,\mu}}[\Om]$ can be taken out of the integral but remains a strongly detuning-dependent prefactor, as evidenced by Fig.~\ref{fig:Dfactor} and the area plots in Fig.~\ref{fig:PSD}.

\section{Ground-state cooling and energy equipartition}\label{app:GS cooling}

\begin{figure}[b]
    \includegraphics[width=\linewidth]{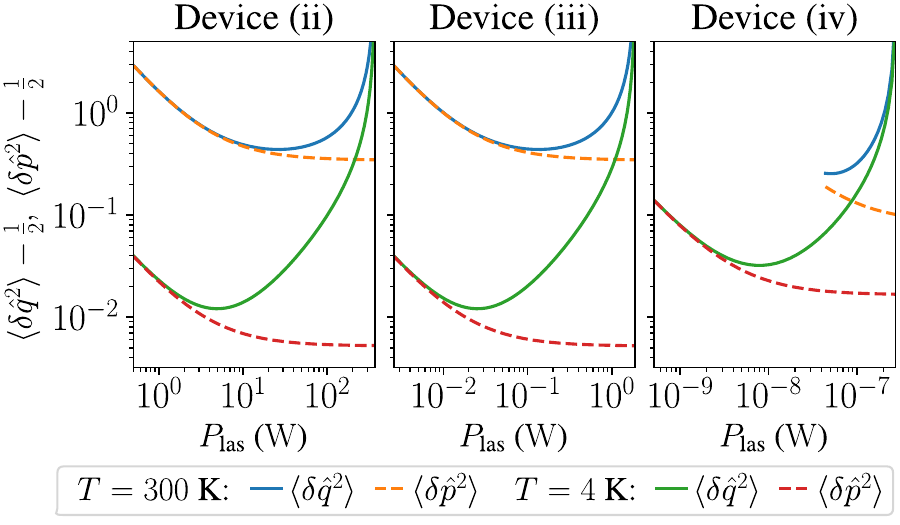}
    \caption{\label{fig:gs cooling}
       Mechanical position and momentum fluctuations, $\mean{\dq^2}$ and $\mean{\deltp^2}$, as a function of laser power for Devices (ii), (iii) and (iv). See Tables \ref{tab:params} and \ref{tab:params2} for the parameters.
    }
\end{figure}

In the main text, we showed that Devices (ii), (iii) and (iv) can reach $\Neff < 1$ (see Figs.~\ref{fig:cooling} and \ref{fig:cooling_device_iv}). This is however not the only requirement to achieve ground-state cooling. The other condition is that energy equipartition, $\mean{\dq^2}\simeq \mean{\deltp^2} \simeq 1/2$, has to be satisfied. We check this second requirement in Fig.~\ref{fig:gs cooling} around the laser powers giving $\Neff < 1$. We see that energy equipartition eventually breaks down when $\Plas$ increases, but on a large range of powers before the one giving the minimum phonon number, we have both $\Neff < 1$ and $\mean{\dq^2}\simeq \mean{\deltp^2}$. This is true both at room temperature (300 K) and at low temperature (4 K), though in the latter case $\mean{\dq^2}$ and $\mean{\deltp^2}$ get closer to 1/2 since the minimum $\Neff$ is smaller in that case.

%

\end{document}